# Multi-Material Topology Optimization of Phononic Crystals for Targeted Elastic Wave Bandgap

Bir Bahadur Chaudhary [a], Wenfeng Li [b], Ting Chen [b], Xinmai Yang [a], Candan Tamerler [a], Bin Ouyang [c], Shiguang Deng [a*]
[a] Department of Mechanical Engineering, University of Kansas, Lawrence, KS, USA
[b] Earth and Environment Science Division, Los Alamos National Laboratory, Los Alamos, NM, USA
[c] Department of Chemical and Biomolecular Engineering, Vanderbilt University, Nashville, TN, USA

## Abstract

Phononic crystals are periodic elastic structures that exhibit frequency bandgaps within which elastic waves cannot propagate. The formation and width of these bandgaps depend on the spatial distribution of material properties within periodic phononic unit cells. In this work, we propose novel topology optimization formulations to achieve multi-material design of phononic crystals for bandgap optimization at targeted frequencies, where we discover the influence of spatially varying properties of base materials on the band structures that block and permit elastic wave propagation. Specifically, we model wave propagation by Floquet-Bloch theory and solve the resulting eigenvalue problem by the finite element method along the boundary of the irreducible Brillouin zone. We first propose a multivariate weight-based method via a novel interpolation scheme that maps the design variables of an arbitrary number of candidate materials into effective density and stiffness for elastic wave-propagation calculations, where bandgaps arise from the contrast in the stiffness and inertia among material constituents. We then develop two formulations for bandgap optimization. The first formulation maximizes the gap between two adjacent frequency bands using a bound-based approach, where we introduce two auxiliary scalar variables as additional design variables and incorporate them directly into the design objective, yielding a linear, differentiable design objective, with the eigenfrequencies entering the formulation through associated constraints. The second formulation places a bandgap over a user-defined target frequency range through a dynamic band-pairing strategy, where we adaptively identify the relevant pair of adjacent bands amid optimization iterations. Towards the end, we validate the bandgap performance of our designs against the transmission loss spectra computed from COMSOL Multiphysics.

## 1. Introduction

Metamaterials are engineered materials whose macroscopic properties derive from a designed internal architecture rather than from the chemical composition of their constituents [1]. Such architecture-driven properties are rare in natural materials [2, 3], and they have enabled unconventional wave-manipulation functionalities across the electromagnetic [4], acoustic [5], and optical [6] domains. Phononic crystals, a class of acoustic metamaterials, represent the elastodynamic counterpart of this family and consist of periodic arrangements of two or more constituent materials that control the propagation of mechanical (elastic) waves, in which waves with certain frequency ranges are forbidden while others propagate freely [7]. This forbidden range is termed the phononic bandgap, and it gives materials their key functional property: a structure excited within the bandgap cannot sustain propagating waves. This property has been exploited in vibration reduction, shock-wave mitigation, seismic protection, noise control, waveguides, and filters [8]. Representative demonstrations include the suppression of flexural vibration in fluid-conveying pipes through periodically designed pipe segments [9] and the reduction of sound power radiated from vibrating bi-material structures [10].

The underlying mechanism of the bandgap is the destructive interference of Bragg-scattered waves produced by a periodic variation of material properties within the unit cell. Early band structure computations for periodic elastic media established that the bandgap width increases as the density contrast between the inclusion and the host material increases [11]; and the first experimental evidence of an absolute bandgap in a solid/solid system was obtained for duralumin cylinders embedded in an epoxy matrix on square and centered-rectangular lattices [12]. Furthermore, for elastic waves, the cermet topology, in which a high-velocity host surrounds isolated low-velocity inclusions, is more favorable for gap formation than the network topology preferred for electromagnetic waves [13]. The dispersion relation is commonly computed using the plane wave expansion method, which requires a Fourier representation of the periodic material properties [14], the finite element method combined with Floquet-Bloch periodic boundary conditions [15], the finite difference time domain method [16], or the transfer matrix method [17]. In these methods, material pairs and inclusion geometries are prescribed a priori, causing the gap to be obtained heuristically, which offers no systematic design principles to realize a desired bandgap. This limitation motivates a systematic, optimization-driven approach for bandgap design.

Topology optimization (TO) is a structural optimization technique that determines the optimal distribution of material within a design domain to extremize a performance objective [18–21], and it has matured along the development of several different algorithms, including density-based methods [22, 23], level-set methods [24, 25], and evolutionary structural optimization [26], which differ in how the material layout is parameterized and updated. Genetic algorithms have also been employed as an optimization strategy for phononic crystal topology design [27, 28]. An application of topology optimization to phononic crystals and bandgap engineering has been reviewed extensively [29–31]. It was established as a systematic route to phononic bandgap design through a two-phase formulation that maximizes the relative gap between adjacent bands over the irreducible Brillouin zone (IBZ) [32].

*Corresponding author
E-mail address: sdeng@ku.edu

Additionally, bandgaps have been maximized in infinite periodic beams modeled by Timoshenko beam theory, where the optimized periodic variation of the cross-sectional properties opens flexural gaps [33]. Beyond maximizing whichever gap emerges between a chosen pair of adjacent bands, a prescribed-bandgap formulation seeks to open and tailor a gap over a user-specified target frequency range. This capability is essential in practice, when the excitation spectrum of the host structure is known in advance. Formulations of this type include a topology optimization approach based on a frequency band constraint together with ipsilateral frequency constraints [34], and a prescribed-bandgap design of bi-material and single-material phononic crystals, in which the region of the unit cell that is not occupied by solid materials is treated as void, and the gap is maximized between user-defined frequency bounds [35].

Classical bandgap optimization studies largely assume a design consisting of one solid, or at most two solid phases. Extending topology optimization to multiple candidate materials rests on a critical challenge: as a discretized material point cannot physically contain several distinct materials simultaneously, an interpolation scheme must map the design variables of all candidate materials onto a single set of effective properties, namely an effective density and an effective elasticity tensor, that enter the governing equations at each point of the design domain. Several such schemes have been developed. An extension of the solid isotropic material with penalization (SIMP) model represents two solid phases together with void using two design variables per finite element and recovers the effective properties through a penalized power-law interpolation whose exponents are selected independently for each phase [36]. For eigenfrequency objectives with multiple anisotropic candidate materials, the effective constitutive matrix has been formed as a penalized weighted sum of the candidate constitutive matrices, while the effective density is interpolated linearly without penalization and an asymmetry is introduced to suppress spurious localized modes [37]. To reduce the large number of design variables and the persistence of intermediate, non-physical material states inherent to penalized weighting, an alternating active-phase algorithm (AAPA) was developed to decompose the multi-phase problem into a sequence of binary sub-problems [38], while a Jacobi-based multi-material model built on the polytopal composite finite element (PCE) technique was developed for incompressible media [39].

From previous studies, we identify three knowledge gaps. First, topology optimization of phononic bandgaps has remained largely binary in terms of material phases. The two-phase formulation becomes considerably less effective, especially when the material contrast is low [32], and the classical fixed-geometry studies naturally confine themselves to prescribed two-phase configurations [11–13]. While recent multi-material work for phononic crystals has begun to move beyond this configuration [40, 41], challenges remain to explicitly treat the difficulties of non-differentiability of repeated eigenfrequencies and the issue of mode-switching that can arise when adjacent bands cross, causing numerical instability during optimization.

Second, existing multi-material studies of adjacent bandgap maximization generally pose the objective in terms of the eigenfrequencies which are directly associated with fixed band indices. They typically require smoothing techniques by aggregation before a gradient-based optimization is applied. This smoothing is effective in practice; however, it may introduce an approximation error and depends on an aggregation parameter that is usually selected empirically.

Third, while few studies [42, 43] have been able to target a prescribed bandgap frequency, most previous formulations maximize the gap between a fixed pair of adjacent bands wherever that gap opens, causing the bandgap location on the frequency range emerging as an outcome of the optimization rather than as a design requirement. Such a prescribed-bandgap formulation appears to be less explored in multi-material design optimization for phononic crystals.

To address these gaps, we propose topology optimization formulations in this work for the design of multi-material phononic crystals with targeted bandgaps. Our key contributions are as follows:

i. To address multi-material interpolation, we propose a multivariate weight-based SIMP interpolation scheme that maps the design variables of an arbitrary number of candidate materials into effective density and effective stiffness in a form suited to wave-propagation objectives in phononic crystals, where bandgap formation depends on the contrasts of material stiffness and inertia.

ii. To address the maximization of the gap between adjacent bands, we adopt a bound formulation [44] approach in which two auxiliary scalar variables serve simultaneously as the objective and as design variables: one is driven upward against the lower band and the other downward against the upper band. The eigenfrequencies enter only through inequality constraints, so the objective is linear in the auxiliary variables and remains differentiable, such that no aggregation approximation nor parameter tuning is required. The pair of bounds identifies which band lies below the gap and which lies above it, so that the target pair remains tracked and correctly ordered throughout the optimization even when neighboring bands cross.

iii. To achieve a bandgap at a target frequency range, we propose a prescribed-bandgap formulation that tailors the gap to a user-defined frequency range. The adjacent-band formulation is augmented with a dynamic band-pairing strategy that adaptively identifies, at each design iteration, the pair of adjacent bands that are most relevant to the prescribed range, so that the resulting gap satisfies the specified frequency requirement.

iv. The effectiveness of our proposed formulations is verified through transmission-loss analysis performed in the frequency domain using finite element models. Furthermore, attenuation bands identified from the simulated transmission-loss spectra validate our bandgap frequencies of the designed crystals.

The remainder of the paper is organized as follows. Section 2 presents the mathematical formulation of the proposed model, followed by the sensitivity analysis in Section 3. Section 4 describes numerical implementation, and Section 5 provides illustrative numerical examples. Section 6 presents numerical validation. Finally, Section 7 concludes the paper.

## 2. Mathematical formulation

### 2.1. Governing equation

Mechanical waves require a material medium for propagation and are classified by the direction of particle displacement relative to the direction of propagation, including longitudinal (pressure or compressional) waves, in which displacement is parallel to propagation, and transverse (shear) waves, in which it is perpendicular. Fluids have negligible shear modulus, and therefore, support only longitudinal waves, whereas solids support both modes. Here, acoustic waves refer to the longitudinal pressure wave, and elastic waves to both wave modes in a solid.

In solid media, elastic waves are generally categorized according to the displacement components involved in the wave motion. Waves involving displacement components confined to a reference plane, conventionally the x-y plane, are termed in-plane modes, whereas waves characterized by displacement components oriented perpendicular to this plane along the *z* direction are referred to as out-of-plane modes [45]. This classification is fundamental in reducing the three-dimensional elastodynamic formulation to a two-dimensional counterpart without loss of physical generality, provided the geometry and loading conditions satisfy the appropriate assumptions.

For the mathematical modeling, we consider in-plane elastic wave propagation in periodic media. For a general three-dimensional, inhomogeneous, isotropic elastic medium, the dynamic equation of motion derived from Newton's second law combined with the constitutive relations of linear elasticity is expressed as ***Equation (1)***:

$$\rho\,\ddot{\mathbf{u}} = \nabla\big((\lambda+\mu)\nabla\cdot\mathbf{u}\big) + \nabla\cdot(\mu\nabla\mathbf{u}), \tag{1}$$

where $\rho$ denotes the material density, $\mathbf{u} = (u, v, w)^T$ represents the displacement vector, and $\lambda$ and $\mu$ are the Lame constants of the elastic medium. For a planar problem, the above equation can be reduced to the two-dimensional elastic wave equations. Under the plane stress assumption, the out-of-plane stress component satisfies $\sigma_z = 0$ and the displacement field is expressed as $\mathbf{u} = (u(x,y,t), v(x,y,t))$, where x and y are the spatial coordinates and 't' is time. Consequently, the governing equations for the in-plane displacement components in the x and y directions are presented in ***Equations (2)*** and ***(3)*** [46]:

$$\rho\frac{\partial^2 u}{\partial t^2} = \mu\left(\frac{\partial^2 u}{\partial x^2}+\frac{\partial^2 u}{\partial y^2}\right) + (\lambda+\mu)\left(\frac{\partial^2 u}{\partial x^2}+\frac{\partial^2 v}{\partial x\,\partial y}\right), \tag{2}$$

$$\rho\frac{\partial^2 v}{\partial t^2} = \mu\left(\frac{\partial^2 v}{\partial x^2}+\frac{\partial^2 v}{\partial y^2}\right) + (\lambda+\mu)\left(\frac{\partial^2 v}{\partial y^2}+\frac{\partial^2 u}{\partial x\,\partial y}\right). \tag{3}$$

### 2.2. Floquet-Bloch wave theory for periodic structures

For periodic media, the wave propagation characteristics can be analyzed using Floquet-Bloch theory [47]. According to this theory, the displacement field in a periodic structure satisfies the Bloch condition, which can be expressed as ***Equation (4)***:

$$\mathbf{u}(\mathbf{y},\mathbf{k}) = \tilde{\mathbf{u}}(\mathbf{y})\,e^{i(\mathbf{k}\cdot\mathbf{y}-\omega t)}, \tag{4}$$

where, $i = \sqrt{-1}$ , $\tilde{\mathbf{u}}$ is a periodic displacement field, $\mathbf{y}$ represents the position vector, and $\mathbf{k}$ is the wave vector across the boundary of the irreducible Brillouin zone $[\Gamma - X - M - \Gamma]$ (see the yellow triangular zone in ***Figure 1***), and $\omega$ is eigen frequency.

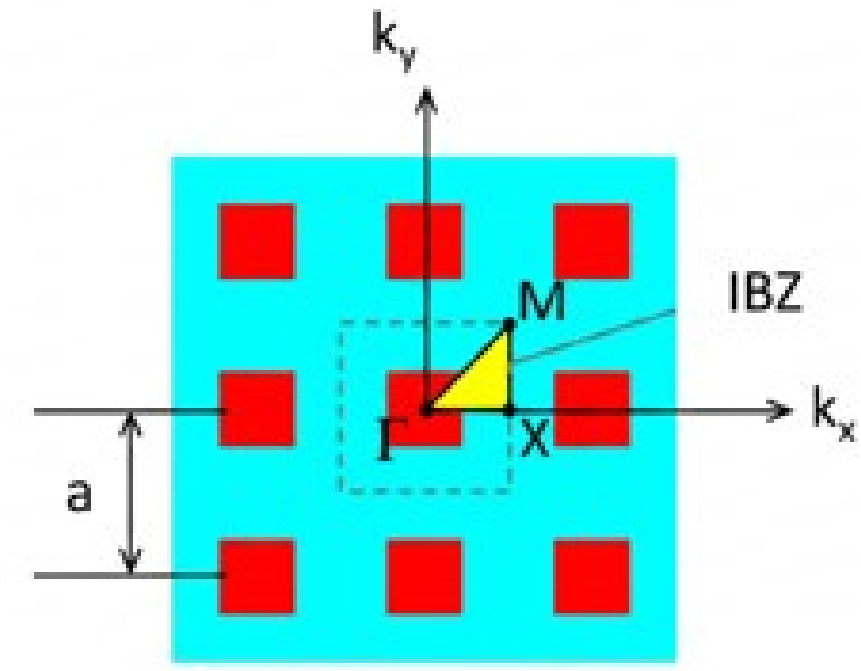

**Figure 1:** An illustration of the irreducible Brillouin zone [48] in a phononic crystal structure consisting of 3-by-3 unit cells, where 'a' indicates the characteristic length of periodic unit cells.

The periodic part of the displacement field satisfies the lattice periodicity condition, which is given by ***Equation (5)***:

$$\tilde{\mathbf{u}}(\mathbf{y}+\mathbf{a},\mathbf{k}) = \tilde{\mathbf{u}}(\mathbf{y},\mathbf{k})\,, \tag{5}$$

where $\mathbf{a} = [a_x, a_y]$ is the lattice vector with a constant.

For a two-dimensional square lattice, the wave vector can be written as ***Equation (6)***:

$$\mathbf{k} = [k_x, k_y],\, k_x = \pm\frac{\pi}{a_x},\, k_y = \pm\frac{\pi}{a_y}. \tag{6}$$

After applying the Bloch boundary conditions and discretizing the governing equations using the finite element method, the wave propagation problem can be written in the following eigenvalue form as given in ***Equation (7)***:

$$(\mathbf{K}(\mathbf{k}) - \omega^2\mathbf{M})\mathbf{u} = 0, \tag{7}$$

where $\mathbf{K}(\mathbf{k})$ is the Bloch stiffness matrix, $\mathbf{M}$ is a mass matrix, $\mathbf{u}$ is the eigenvector (mode shape). The eigenvectors are typically mass-normalized, which leads to the orthogonality condition as given in ***Equation (8)***:

$$\mathbf{v}_i^T\mathbf{M}\mathbf{v}_j = \begin{cases}0 & i \neq j\\ 1 & i = j\end{cases}, \tag{8}$$

where $\mathbf{v}_i$ and $\mathbf{v}_j$ denote the eigenvectors corresponding to the i-th and j-th eigenfrequencies. Since IBZ is symmetric to the whole unit cell, only the design variables inside the triangular IBZ are considered during optimization [49]. These reduced design variables are later mapped back to the original full domain using symmetric mapping (see mapping matrix in ***Appendix B***).

### 2.3. Multi-material topology optimization

We discretize the design domain into $n_e = n_{elx} \times n_{ely}$ four-node quadrilateral elements, where $n_{elx}$ and $n_{ely}$ denote the number of elements in the horizontal and vertical directions, respectively. We start with extending classical SIMP that typically works for bi-material designs to a multivariate weight-based formulation for $n_{Mat}$ candidate materials by introducing $n_{Mat} - 1$ design fields (f) per element. For example, with a three-material phononic crystal, it gives two fields, so the design variable vector for element 'e' becomes $x_e = [x_e^{(1)}, x_e^{(2)}]$ with $x_e^{(1)},\ x_e^{(2)}\ \in [0,1]$, where (1) and (2) in superscripts indicate the design field. For the three-material case as shown in ***Figure 2***, if $x_e^{(1)} = 0$, we select the first material (M1); if $x_e^{(1)} = 1$ and $x_e^{(2)} =$ 0, we select the second material (M2); and if $x_e^{(1)} = 1$ and $x_e^{(2)} = 1$, we select the third material (M3).

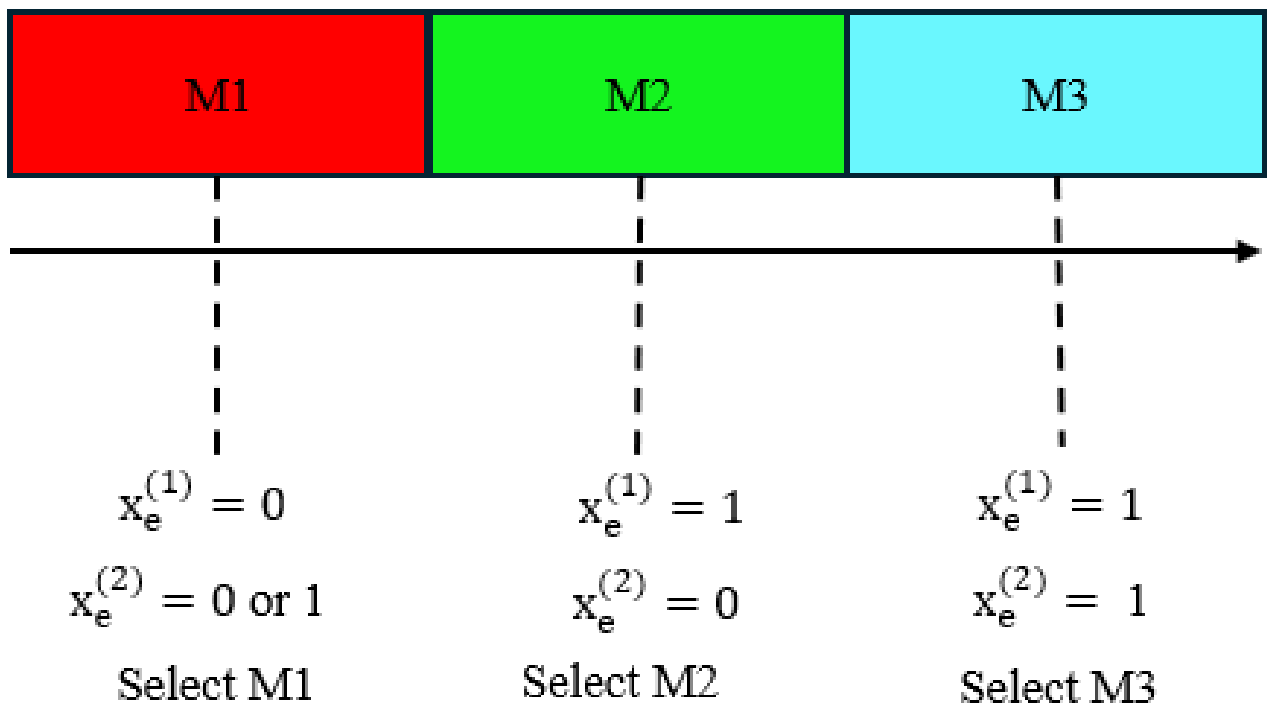

**Figure 2:** Proposed novel multivariate weight-based multi-material interpolation scheme

To remove the checkerboard pattern and mesh dependency, we filter the design variables with a density filter [50] and obtain the filtered densities ($\tilde{x}_e = [\tilde{x}_e^{(1)}, \tilde{x}_e^{(2)}]$) from ***Equation (9)***:

$$\tilde{x}_e^{(f)} = \frac{\sum_{j \in N_e} w(x_j)\, v_j\, x_j^{(f)}}{\sum_{j \in N_e} w(x_j)\, v_j}, \tag{9}$$

where $v_j$ and $x_j^{(f)}$ represent the volume and density per design field of element j, $N_e$ is the neighborhood set of elements lying within the filter domain for element e, $w(\mathbf{x})$ is the weighting function, which is defined as $w(x_j) = r_{min} - \|x_j - x_i\|$, with filter radius $r_{min}$ and, $x_i$ and $x_j$ contain the central coordinates of the design cells i and j, respectively. We then reduce intermediate densities and obtain a black-and-white design by projecting the filtered densities through the smoothed Heaviside function [50] as given by ***Equation (10)***:

$$\phi_e^{(f)} = \frac{\tanh(\beta_f \eta) + \tanh(\beta_f(\tilde{x}_e^{(f)} - \eta))}{\tanh(\beta_f \eta) + \tanh(\beta_f(1 - \eta))}, \tag{10}$$

which yields the projected fields ($\phi_e = [\phi_e^{(f)}, \phi_e^{(f)}]$). Here, $\eta$ is the projection threshold and $\beta_f$ controls the sharpness of the projection. We increase $\beta_f$ gradually over the course of the optimization. Differentiating ***Equation (10)*** with respect to the filtered density gives the sensitivity of the projection, which is shown in ***Equation (11)***:

$$\frac{\partial \phi_e^{(f)}}{\partial \tilde{x}_e^{(f)}} = \frac{\beta_f\left(1 - \tanh^2(\beta_f(\tilde{x}_e^{(f)} - \eta))\right)}{\tanh(\beta_f \eta) + \tanh(\beta_f(1 - \eta))}, \tag{11}$$

which we use in the chain rule (see ***Section 3***), when assembling the design sensitivities.

We build multi-material interpolation on the multivariate weight-based SIMP scheme, in which each material draws a weight from the projected design fields. For a mesh of $n_e$ elements, we define the unnormalized weight in vector form for the first material as given by ***Equation (12)***:

$$\mathbf{w}_1^{un} = \boldsymbol{\phi}_1^r, \tag{12}$$

where r denotes the penalization parameter and $\phi_1$ represents the projected design field associated with the first material. Similarly, the unnormalized weight for $2 \le m \le n_{Mat} - 1$ materials are given by ***Equation (13)***:

$$\mathbf{w}_m^{un} = \left(\prod_{j=1}^{m-1} (1 - \boldsymbol{\phi}_j^r)\right) \boldsymbol{\phi}_m^r. \tag{13}$$

The unnormalized weight for $n_{Mat}^{th}$ material is given by ***Equation (14)***:

$$\mathbf{w}_{n_{Mat}}^{un} = \prod_{j=1}^{n_{Mat}-1} (1 - \boldsymbol{\phi}_j^r). \tag{14}$$

Finally, we normalized all weights as per ***Equation (15)***:

$$\mathbf{w}_m = \frac{\mathbf{w}_m^{un}}{\sum_{m=1}^{n_{Mat}} \mathbf{w}_m^{un}}, \tag{15}$$

so that they satisfy partition of unity condition $\sum_{m=1}^{n_{Mat}} \mathbf{w}_m = 1$.

### 2.4. Finite element formulation

Under the plane stress assumption, the thickness of our design domain is much smaller than its in-plane dimensions. The two-dimensional finite element solution can be formulated with the constitutive matrix defined as in ***Equation (16)***. We assemble the element stiffness and mass matrices into global matrices, which leads to the eigenvalue problem in ***Equation (7)***. Its solution yields the eigenvalues and eigenvectors, where we mass-normalize the eigenvectors according to ***Equation (8)***. The plane stress constitutive matrix for 'm' material is:

$$\mathbf{D_m} = \frac{E_m}{1 - {\nu_m}^2} \begin{bmatrix} 1 & \nu_m & 0 \\ \nu_m & 1 & 0 \\ 0 & 0 & \frac{1 - \nu_m}{2} \end{bmatrix}. \tag{16}$$

Element stiffness and mass matrices for the 'm' material are calculated as per ***Equation (17)***:

$$\mathbf{K_e^{(m)}} = \int_{\Omega_e} \mathbf{B}^T \mathbf{D}_m \mathbf{B}\, d\Omega, \qquad \mathbf{M_e^{(m)}} = \int_{\Omega_e} \rho_m N^T N\, d\Omega, \tag{17}$$

where B is the strain-displacement matrix for the candidate material, and N is the shape function. Elemental stiffness and mass matrices with the weight vector are as per ***Equation (18)***:

$$\mathbf{K_e} = \sum_{m=1}^{n_{Mat}} w_{m,e} \mathbf{K}_e^{(m)}, \mathbf{M_e} = \sum_{m=1}^{n_{Mat}} w_{m,e} \mathbf{M}_e^{(m)}. \tag{18}$$

Global stiffness and mass matrices are given by ***Equation (19)***:

$$\mathbf{K_g} = \sum_{e=1}^{n_e} \mathbf{A}_e^T \mathbf{K}_e \mathbf{A}_e, \mathbf{M_g} = \sum_{e=1}^{n_e} \mathbf{A}_e^T \mathbf{M}_e \mathbf{A}_e, \tag{19}$$

where, $\mathbf{A}_e$ represents the matrix to map the degrees of freedom (DOF) from elements to the global. To ensure the symmetry of the stiffness and mass matrices, the assembled matrices are symmetrized as per ***Equation (20)***:

$$\mathbf{K_g} = \frac{1}{2}(\mathbf{K_g} + \mathbf{K_g}^T), \mathbf{M_g} = \frac{1}{2}(\mathbf{M_g} + \mathbf{M_g}^T). \tag{20}$$

### 2.5. Bandgap maximization of adjacent eigenfrequencies

In this section, we formulate the topology optimization problem for maximizing the bandgap between two adjacent eigenfrequencies of a phononic crystal. The design objective is to widen the frequency range between the $j^{th}$ and $(j+1)^{th}$ eigenfrequencies, so that elastic waves in this range cannot propagate through the periodic medium. The optimization formulation is given by ***Equation (21)***:

$$
\begin{aligned}
&\max_{\mathbf{x}_e} \quad \min_{\mathbf{k}} \ \omega_{j+1}(\mathbf{k}, x_e) - \max_{\mathbf{k}} \ \omega_j(\mathbf{k}, x_e) \qquad (21)\\
&\text{s.t.} \quad [\mathbf{K}(\mathbf{k},\mathbf{x}) - \omega^2\mathbf{M}(\mathbf{x})]\mathbf{u} = \mathbf{0}\\
&\frac{v_m}{v_m^{target}} - 1 \le 0, m = 1, \dots, n_{Mat}\\
&\mathbf{k} \in [\Gamma - X - M - \Gamma]\\
&x_e = [x_e^{(f=1)}, x_e^{(f=2)}, \dots], \quad 0 \le x_e^{(f)} \le 1,\\
&f = 1, \dots, n_{mat} - 1, \quad e = 1, \dots, n_e,
\end{aligned}
$$

where, $\omega_j$ is the jth eigenfrequency, and the volume of the candidate material during optimization is calculated as per ***Equation (22)***:

$$
v_m = \frac{1}{n_e}\sum_{e=1}^{n_e} w_m\,(e); \quad m = 1, \dots, n_{Mat}. \qquad (22)
$$

This optimization problem in ***Equation (21)*** is a min-max formulation that is non-differentiable, and therefore, difficult to handle by gradient-based optimization. To overcome this difficulty, we used a bound formulation [51], [52] by employing auxiliary scalar variables to represent the lower and upper bounds of the bandgap. In this approach, the scalar variables serve as both design variables and the objective, enabling the bandgap maximization to be reformulated as a differentiable formulation. An additional advantage of the bound formulation is that it effectively avoids *mode switching* during the optimization process, since the objective depends only on the auxiliary bound variables while the eigenvalues enter through constraints imposed on all modes of each band, so that a crossing of two eigenvalues leaves the objective and its sensitivities unchanged. Furthermore, the max and min operators appearing in the bandgap definition are approximated using smooth p-norm aggregation functions. The maximum and minimum operators for the lower and upper bands are approximated using the p-norm function given by ***Equation (23)*** [53]. The parameters $\lambda_{ref,lower}$ and $\lambda_{ref,upper}$ are introduced as scaling factors to improve numerical stability. In this formulation, these scaling factors are taken as $\omega_{j_{lower}}$ and $\omega_{j_{lower}+1}$, respectively, during the first iteration of the optimization loop, where $j_{lower}$ and $j_{lower}+1$ are the eigenmodes between which a band gap is created.

$$
\omega_p^{(j)} = \left(\sum_{n=1}^{N_k} \omega_j\,(\mathbf{k}(n))^p\right)^{1/p}, \qquad (23)
$$

$$
\omega_{pu}^{(j)} = \left(\sum_{n=1}^{N_k} \omega_j\,(\mathbf{k}(n))^{-p}\right)^{-1/p},
$$

where $N_k$ is the total number of wave vectors, and p is the p-norm parameter. Our formulation becomes:

$$
\begin{aligned}
&\max_{x_e,\lambda_1,\lambda_2} \quad \lambda_2 - \lambda_1 \qquad (24)\\
&\text{s.t.} \quad [\mathbf{K}(\mathbf{k},\mathbf{x}) - \omega^2\mathbf{M}(\mathbf{x})]\mathbf{u} = 0,\\
&g^{(1)}: \frac{\omega_p^{(j)}}{\lambda_{ref,lower}} - \lambda_1 \le 0,\\
&j = 1, \dots, j_{lower},\\
&g^{(2)}: \lambda_2 - \frac{\omega_p^{(j)}}{\lambda_{ref,upper}} \le 0,\\
&j = j_{lower} + 1, \dots, j_{upper},\\
&g^{(3)}: \frac{v_m}{v_m^{target}} - 1 \le 0,\\
&m = 1, \dots, n_{Mat},\\
&\mathbf{k} \in [\Gamma - X - M - \Gamma],\\
&x_e = [x_e^{(f=1)}, x_e^{(f=2)}, \dots], 0 \le x_e^{(f)} \le 1,\\
&f = 1, \dots, n_{Mat} - 1,\\
&0 \le \lambda_1 \le 1, \quad 0 \le \lambda_2 \le 1.
\end{aligned}
$$

Here, the objective is to maximize the frequency gap between two adjacent eigenmodes, and scalar variables $\lambda_1$ and $\lambda_2$ are used in the objective function as $\lambda_2 - \lambda_1$. They will also act as design variables as shown in ***Equation (24)***. The constraint functions are eigenfrequency constraints for the lower mode ($g^{(1)}$) and upper mode ($g^{(2)}$), and volume fraction constraints ($g^{(3)}$) for each material. Their sensitivity calculation is given in subsequent sections. The total number of design variables for this TO formulation is given in ***Equation (25)***:

$$
\mathbf{x} = \left\{\left\{\left\{x_e^{(1)}\right\}_{e=1}^{n_e}, \left\{x_e^{(2)}\right\}_{e=1}^{n_e}\right\}, \lambda_1, \lambda_2\right\}, \qquad (25)
$$

where, for instance, $x_e^{(1)}$ is the design variable for element (e) for the first type of material, as the material selection as $f = 1$.

### 2.6. Bandgap maximization between targeted frequencies

In this section, we formulate the topology optimization problem for maximizing a bandgap over a prescribed frequency range. Unlike the previous formulation, where the gap location is an outcome of the optimization, we specifiy the lower and upper bounds of the target range a priori, and the design is driven so that no eigenfrequency lies in this range. The optimization formulation is given by ***Equation (26)***:

$$
\begin{aligned}
&\max_{x_e} \quad \frac{\omega_p^{upper} - \omega_p^{lower}}{\omega_{low,target}} \qquad (26)\\
&\text{s.t.} \quad g^{(1)}: \frac{\omega_p^{lower}}{\omega_{low,target}} - 1 \le 0\\
&g^{(2)}: \frac{\omega_{upp,target}}{\omega_p^{upper}} - 1 \le 0\\
&g^{(3)}: \frac{V_m}{V_m^{target}} - 1 \le 0, m = 1, \dots, n_{mat}\\
&[\mathbf{K}(\mathbf{k},\mathbf{x}) - \omega^2\mathbf{M}(\mathbf{x})]\mathbf{u} = 0\\
&\mathbf{k} \in [\Gamma - X - M - \Gamma]\\
&x_e = [x_e^{(f=1)}, x_e^{(f=2)}, \dots], 0 \le x_e^{(f)} \le 1,\\
&f = 1, \dots, n_{mat} - 1, \quad e = 1, \dots, n_e.
\end{aligned}
$$

Lower band (smoothed maximum across k-points) and upper band (smoothed minimum across k-points) are given by ***Equation (27)***:

$$\omega_p^{lower} = \left(\sum_{n=1}^{N_k} \omega_j\,(\mathbf{k}(n))^p\right)^{1/p}, \quad \omega_p^{upper} = \left(\sum_{n=1}^{N_k} \omega_j\,(\mathbf{k}(n))^{-p}\right)^{-1/p}. \tag{27}$$

This optimization formulation, given by ***Equation (26)***, aims to create a bandgap within the prescribed frequency range of $[\omega_{low,target}, \omega_{upp,target}]$. Our method maximizes the bandgap between consecutive eigenfrequencies via a *dynamic band-pair selection strategy*. It starts with maximizing a bandgap between the first and second eigenfrequencies. If the resulting bandgap does not satisfy the prescribed target bandgap range, our algorithm proceeds to maximize the bandgap between the second and third eigenfrequencies, then between the third and fourth eigenfrequencies, and so on. The sensitivity analysis required for this topology optimization formulation is presented in the subsequent sections. The total set of design variables for this formulation is given by ***Equation (28)***:

$$\mathbf{x} = \left\{\{x_e^{(1)}\}_{e=1}^{n_e}, \{x_e^{(2)}\}_{e=1}^{n_e}\right\}. \tag{28}$$

Here, $\mathbf{x} \in \mathbb{R}^{n_e \times f}$; where $n_e$ and f denote the number of finite elements and design fields, respectively.

## 3. Sensitivity Calculations

### 3.1. Gradient of weights

To evaluate the sensitivities of the design objective and constraints, we first need to calculate the gradient of weights associated with the multivariate weight-based SIMP interpolation. Here, f denotes a variable for the number of design fields, which is $1 \le f \le n_{Mat} - 1$, and m denotes a variable for the number of materials, $2 \le m \le n_{Mat}$. The gradient of the unnormalized weight is calculated in ***Equation (29)***. The detailed calculation of the gradient of weights (w) for each material is given in ***Appendix A.1***.

$$\frac{\partial \mathbf{w}_m}{\partial \boldsymbol{\phi}_f} = \frac{\left(\frac{\partial \mathbf{w}_m^{un}}{\partial \boldsymbol{\phi}_f}\right)\sum_{m=1}^{n_{Mat}} \mathbf{w}_m^{un} - \mathbf{w}_m^{un}\left(\sum_{m=1}^{n_{Mat}} \frac{\partial \mathbf{w}_m^{un}}{\partial \boldsymbol{\phi}_f}\right)}{\left(\sum_{m=1}^{n_{Mat}} \mathbf{w}_m^{un}\right)^2}. \tag{29}$$

### 3.2. Sensitivity for bandgap maximization between adjacent eigenfrequencies

The design problem is a multi-constrained, gradient-based optimization, which requires the sensitivities of the objective as well as of each constraint, where sensitivity fields must be derived for every constraint before the constraints are aggregated in the optimization framework [54, 55] as below.

#### 3.2.1 Sensitivity of objective function

The sensitivity of the objective function is calculated with respect to both density design variables and auxiliary variables, which is given in ***Equation (30)***. Detailed calculation of the sensitivity of the objective function is given in ***Appendix A.2 (i)***.

$$\frac{\partial f_0}{\partial x_e} = [0, \dots, 0,\ 1,\ -1]^T. \tag{30}$$

As the objective function is a function of the auxiliary variables alone, its derivatives with respect to the density variables are zero, and its derivatives with respect to the auxiliary variables are 1 and -1, respectively.

#### 3.2.2 Sensitivity of lower band constraints

Since the design variable follows this sequence $\mathbf{x} \rightarrow \tilde{\mathbf{x}} \rightarrow \boldsymbol{\phi} \rightarrow \mathbf{w} \rightarrow \mathbf{K}, \mathbf{M},$ the chain rule is applied to calculate sensitivities. After calculating sensitivity with respect to the projected design variables, density filtering is applied using the chain rule. Sensitivity is mapped with respect to the reduced density variables by IBZ symmetry. Detailed sensitivity calculations are given in ***Appendix A.2 (ii)***.

For band indices $j = 1, \dots, j_{lower}$, sensitivity is given by ***Equation (31)***:

$$\frac{\partial g_j^{(1)}}{\partial x_e} = \frac{1}{\lambda_{ref}} \frac{\partial \omega_p^{(j)}}{\partial x_e}, \tag{31}$$

where,

$$\frac{\partial \omega_p^{(j)}}{\partial x_e} = \omega_p^{(j)\,(1-p)} \sum_{n=1}^{N_k} \omega_{\mathbf{k}(n),j}^{(p-1)} \frac{\partial \omega_{\mathbf{k}(n),j}}{\partial x_e}. \tag{32}$$

***Equation (32)*** is further simplified as ***Equation (33)***:

$$\frac{\partial \omega_{\mathbf{k},j}}{\partial x_e} = \frac{1}{2\omega_{\mathbf{k},j}} \mathbf{u}_{\mathbf{k},j}^T \left(\frac{\partial \mathbf{K}_e}{\partial x_e} - \omega_{\mathbf{k},j}^2 \frac{\partial \mathbf{M}_e}{\partial x_e}\right) \mathbf{u}_{\mathbf{k},j}. \tag{33}$$

After calculating sensitivity with respect to the projected design variables, density filtering is applied using the chain rule. Density filtering of sensitivity is given by ***Equation (34)***:

$$\widetilde{\frac{\partial g_j^{(1)}}{\partial \mathbf{x}}} = \mathbf{H}\left(\frac{\frac{\partial g_j^{(1)}}{\partial \mathbf{x}}}{\mathbf{H_s}}\right). \tag{34}$$

Here, the tilde symbol over $\frac{\partial g_j^{(1)}}{\partial \mathbf{x}}$ indicates filtering of sensitivity. Mapping to reduced variables, $\mathbf{x_{red}}$ by IBZ symmetry is given in ***Equation (35)***:

$$\frac{\partial g_j^{(1)}}{\partial \mathbf{x_{red}}} = \frac{1}{\lambda_{ref}} \mathbf{M_{MAP}}^T \widetilde{\frac{\partial g_j^{(1)}}{\partial \mathbf{x}}}. \tag{35}$$

The sensitivity w.r.t. the first auxiliary variable $\lambda_1$ is given by ***Equation (36)***:

$$\frac{\partial g_j^{(1)}}{\partial \lambda_1} = -1. \tag{36}$$

Since the lower-band constraints do not depend on the second auxiliary variable $\lambda_2$, its derivative with respect to that variable vanishes. The complete sensitivity is then obtained by assembling the derivatives with respect to all design variables. Specifically, the density variables with respect to the auxiliary variables are given in ***Equation (37)***:

$$\frac{\partial g_j^{(1)}}{\partial \mathbf{x_{mma}}} = \left[\frac{\partial g_j^{(1)}}{\partial \mathbf{x_{red}}}, \quad \frac{\partial g_j^{(1)}}{\partial \lambda_1}, \quad 0\right]. \tag{37}$$

#### 3.2.3 Sensitivity of upper band constraints

For upper bands $u = 1, \dots, n_{upper}$, with $j = j_{lower} + u$, sensitivity is given in ***Equation (38)***. This sensitivity is expanded by the chain rule for the Heaviside projection and density filtering. Then, the sensitivity is reduced in size using the symmetry mapping matrix $\mathbf{M_{MAP}}$. The derivatives of the upper-band constraints with respect to the two auxiliary variables are evaluated. The derivative with respect to the first auxiliary variable is zero. The contributions from all design variables are then assembled, where the detailed calculation of sensitivity is provided in ***Appendix A.2 (iii)***.

$$\frac{\partial g_u^{(2)}}{\partial x_e} = -\frac{1}{\lambda_{refu}} \frac{\partial \omega_{pu}^{(u)}}{\partial x_e}. \tag{38}$$

#### 3.2.4 Sensitivity of volume constraints

For each material, the sensitivity of the volume constraint follows from the corresponding volume fraction in ***Equation (39)***, expanded through the chain rule for the Heaviside projection and the density filter, and reduced in dimension using the symmetry mapping matrix $\mathbf{M_{MAP}}$. Since volume constraints are independent of the auxiliary variables, the associated derivatives vanish. By assembling the contributions from all design variables, we provide the full derivation in ***Appendix A.2(iv)***. For material $m = 1, \dots, n_{Mat}$

$$\frac{\partial g_m^{(3)}}{\partial x_e} = \frac{1}{v_m^{target}} \frac{\partial v_m}{\partial x_e}. \tag{39}$$

### 3.3. Sensitivity of bandgap maximization between targeted frequencies

In this formulation, there is no auxiliary variable. Sensitivity is, therefore, calculated only with respect to the pseudo-density design variable.

#### 3.3.1 Sensitivity of objective function

Sensitivity of the objective function is calculated as per ***Equation (40)***. Detailed calculations are in ***Appendix A.3 (i)***.

$$\frac{\partial f_0}{\partial x_e} = -\frac{1}{\omega_{low,\ target}} \left( \frac{\partial \omega_p^{upper}}{\partial x_e} - \frac{\partial \omega_p^{lower}}{\partial x_e} \right). \tag{40}$$

#### 3.3.2 Sensitivity of lower band constraints

For a lower band index $j = j_{lower}$, sensitivity is given by ***Equation (41)***, and calculations are provided in ***Appendix A.3 (ii)***.

$$\frac{\partial g_1}{\partial x_e} = \frac{\omega_p^{lower\ (1-p)}}{\omega_{low,\ target}} \sum_{n=1}^{N_k} \omega_{\mathbf{k}(n),j}^{p-1} \frac{\partial \omega_{\mathbf{k}(n),j}}{\partial x_e}. \tag{41}$$

#### 3.3.3 Sensitivity of upper band constraints

For the upper band index $u = j_{upper}$, sensitivity is given by ***Equation (42)***, and the calculation is given in ***Appendix A.3 (iii)***.

$$\frac{\partial g_2}{\partial x_e} = -\omega_{upp} \left( \omega_p^{upper} \right)^{p-1} \sum_{n=1}^{N_k} \omega_{\mathbf{k}(n),j+1}^{-p-1} \frac{\partial \omega_{\mathbf{k}(n),j+1}}{\partial x_e}. \tag{42}$$

#### 3.3.4 Sensitivity of volume constraints

The sensitivity to volume constraints in topology optimization for maximum bandgap at targeted frequency range is the same as in topology optimization for adjacent eigenfrequencies bandgap maximization.

## 4. Numerical Implementation

We present the numerical implementation for band gap formation due to Bragg scattering, that is, from the destructive interference of waves scattered by the periodic arrangement of unit cells. Accordingly, the geometries and material distributions adopted in the following examples are configured to exploit impedance contrast and periodicity, and the computed band gaps are located at wavelengths comparable to the unit cell size. We note that local resonance, which arises when a high-density inclusion embedded in a soft matrix acts as an equivalent mass-spring system to generate bandgaps at wavelengths much larger than the unit cell, is not included in this work.

### 4.1. Material properties

The materials listed in ***Table 1*** are selected for our multi-material phononic crystal bandgap optimization. Alongside the elastic modulus (E), density (ρ), and Poisson's ratio (ν), it includes Lame constants, wave speeds, and characteristic impedances, where the Lame constants are calculated by ***Equation (43)***:

$$\lambda = \frac{\nu E}{1 - \nu^2}, \qquad \mu = \frac{E}{2(1 + \nu)}. \tag{43}$$

The longitudinal and shear wave speeds are calculated by ***Equation (44)***:

$$c_P = \sqrt{\frac{E}{\rho(1 - \nu^2)}}, \qquad c_s = \sqrt{\frac{\mu}{\rho}}, \tag{44}$$

and the longitudinal and shear impedances are calculated by ***Equation (45)***:

$$Z_P = \rho c_P, \qquad Z_s = \rho c_s. \tag{45}$$

These impedances govern the reflection and transmission of in-plane elastic waves at internal interfaces, and, therefore, the formation of bandgaps. The gaps obtained belong to the Bragg type. Since the shear-wave velocity is lower than the longitudinal-wave velocity for all constituent materials considered in this work, the shear-dominated (transverse) branches generally occur at lower frequencies in the dispersion diagram. The reflection strength at each interface is determined by the impedance mismatch between adjacent phases, while the gap frequency depends on the lattice constant and the effective wave speed in the multi-material phononic crystal. We note that a distinction from the previous study [56] is that our constituents are not ordered monotonically in elasticity and density. Specifically, conventional multi-material formulations [56] typically assume a material sequence that increases monotonically in stiffness and density. Our multi-material selection method deliberately relaxes this assumption, admitting an arbitrary ordering in which stiffness and density vary independently.

**Table 1:** Material properties

| Material | Modulus of Elasticity, E, (GPa) | Density, ρ, ($kg/m^3$) | Poisson's ratio, ν | Lame constant λ, (GPa) | Lame constant, μ, (GPa) | Longitudinal wave speed, $c_P$, ($\times 10^3$ m/s) | Longitudinal impedance, $Z_P$, ($\times 10^6$ $kg/m^2.s$) | Shear wave speed, $c_S$, ($\times 10^3$ m/s) | Shear impedance, $Z_S$, ($\times 10^6$ $kg/m^2.s$) |
|---|---|---|---|---|---|---|---|---|---|
| Lead (Pb) | 17 | 11344 | 0.42 | 8.67 | 5.99 | 1.35 | 1.53 | 0.73 | 8.24 |
| Epoxy (Ep) | 2.4 | 1200 | 0.35 | 0.96 | 0.89 | 1.51 | 1.81 | 0.86 | 1.03 |
| Brass (Br) | 97 | 8490 | 0.31 | 3.33 | 3.70 | 3.56 | 3.02 | 2.09 | 1.77 |
| Aluminum (Al) | 70 | 2700 | 0.35 | 2.79 | 2.59 | 5.44 | 1.47 | 3.10 | 8.37 |

### 4.2. Proposed algorithm

Our algorithm is demonstrated below, which adopts the MMA optimizer [57], SIMP-based material penalization [58, 59], and material interpolation scheme [60]. We adopt the following parameters for our numerical studies:

- **Penalization power:** In both formulations, the penalization exponents for stiffness and mass interpolation are initialized at unity. A continuation scheme is then applied, in which the stiffness penalization exponent r is increased gradually during the optimization, and the mass penalization is left unpenalized throughout. For the first formulation, r is incremented by 0.25 every 25 iterations once the 100th iteration has been passed, subject to an upper bound of 3. For the second formulation, the increment is reduced to 0.10 and applied every 25 iterations up to the same maximum of 3. Continuation is initiated at the 100th iteration for the two- and four-material cases and at the 50th iteration for the three-material case. These values were chosen because they produce optimized topologies containing few gray elements.
- **Initial design variable:** For both formulations, the initial value of design variables for each finite element is set to 0.2. For example, the initial value for two-material is $x_e = 0.2$, for three-material is $x_e = [0.2, 0.2]$, and for four-material $x_e = [0.2, 0.2, 0.2]$. The design variables are initialized uniformly at $x_e = 0.2$. This value is not selected to coincide with the prescribed volume fraction; rather, it is adopted as an unbiased initialization.
- **Scalar variable:** For the first formulation, $\lambda_1$ and $\lambda_2$ are initialized with 0.4 and 0.6, respectively, with bounds [0, 1].
- **Filter radius:** In both formulations, the radius of the density filter is set to 2.5 for the two-material case and to 5 for the three- and four-material cases.
- **Projection parameter:** The projection parameter is initialized at 1 for the bi-material case and at 3 for the tri- and quad-material cases in the first formulation, and at 1 for all three examples in the second formulation.
- **Projection threshold value:** The threshold value of the projection function is taken as 0.5.
- **P-norm parameter:** The parameter used in the p-norm approximation for the max-min functions is set to 30.
- **Number of wave vectors:** The number of wave vectors sampled along each side of the irreducible Brillouin zone is 20.
- **Number of frequency bands:** The number of bands considered in our analysis is 9.
- **Stopping criteria:** The optimization is terminated once the bandgap width stagnates over successive iterations and the continuation parameter has reached its maximum value.

**Proposed Algorithm:** Multi-material topology optimization of phononic bandgaps

**Inputs:** number of candidate material $n_m$; material data $\{E_m, \nu_m, \rho_m\}$, $m = 1, \ldots, n_m$; mesh data; number of bands to compute $n_b$; maximum number of iterations $N_{max}$.
**Output:** optimized design variables $\mathbf{x}$.
**Initialization:** $\mathbf{x} \leftarrow \mathbf{x_0}$; loop $\leftarrow 0$;
$r_{min} \leftarrow r_{min,0}$, $\beta_f \leftarrow \beta_{f0}$, $p \leftarrow p_0$ ▷ filter radius, projection and p-norm parameters
assemble symmetry mapping matrix $\mathbf{M}_{MAP}$
build wave-vector path $\{\mathbf{k}_j\}$, $j = 1, \ldots, n_k$, along the IBZ boundary ▷ Floquet-Bloch BC
initialize MMA parameters ($\mathbf{x}^{loop-1}$, $\mathbf{x}^{loop-2}$, $\mathbf{L}$, $\mathbf{U}$)
while loop < $N^{max}$ and not converged do
  $\mathbf{x} \leftarrow \{x_e^{(f=1)}, x_e^{(f=2)}, \ldots, x_e^{(f=n_m)}\}$ ▷ element-wise design fields
  $\tilde{\mathbf{x}} \leftarrow$ Filter ($\mathbf{x}$, $r_{min}$)
  $\boldsymbol{\phi} \leftarrow$ Projection ($\tilde{\mathbf{x}}$, $\beta_f$, η)
  $\mathbf{w} \leftarrow$ weight ($\boldsymbol{\phi}$)
  $\mathbf{K}(\mathbf{k}_j)$, $\mathbf{M}(\mathbf{k}_j) \leftarrow \mathbf{T}^T\mathbf{K_0}(\mathbf{w})\mathbf{T}$, $\mathbf{T}^T\mathbf{M_0}(\mathbf{w})\mathbf{T}$ ▷ apply Floquet-Bloch BC
  solve $[\mathbf{K}(\mathbf{k}_j) - \omega_n^2(\mathbf{k}_j)\,\mathbf{M}(\mathbf{k}_j)]\,\mathbf{u}_n = \mathbf{0}$, $n = 1, \ldots, n_b$, $j = 1, \ldots, n_k$
  evaluate objective and sensitivity: $f_0(\mathbf{x})$, $\frac{\partial f0}{\partial \mathbf{x}}$

```
        evaluate constraints and sensitivities: f(x), ∂f/∂x
        x_red ← M_MAP^T x                                          ▷ reduced design variables
        x_red,mma ← MMA (x_red, f0(x_red), f(x_red), ∂f0/∂x_red, ∂f/∂x_red)
        x ← M_MAP x_red,mma                                        ▷ map back to the full design space
        if loop > 100 and mod (loop, 25) = 0 then
            r ← min (r + 0.25, 3)                                  ▷ stiffness penalization continuation
        end
        if mod (loop, 25) = 0 then
            βf ← min (2βf, 300)                                    ▷ projection continuation
        end
        if loop = Nmax & f0^(loop-1) ≈ f0^loop then
            converged ← true                                       ▷ check convergence
        end
        loop ← loop + 1
    end while
    return x
```

## 5. Numerical Examples

In this study, we consider the dimensions of the two-dimensional unit cell of a phononic crystal as 0.1 m × 0.1 m, where a schematic illustration of the unit cell is shown in ***Figure 3***. Since the unit cell serves as the fundamental building block of the phononic crystal, infinite periodic boundary conditions are applied on all four boundaries to represent an infinitely periodic structure. The design domain is discretized into 100 × 100 finite elements. For the finite element analysis, a plane stress condition is assumed. The domain is discretized using four-node bilinear quadrilateral elements, and the element matrices are evaluated using four-point Gauss quadrature for numerical integration. The material properties used in this study are summarized in ***Table 1***. The volume fractions of all candidate materials are set to be equal, imposing an equal-material distribution within the design domain. This constraint ensures that every candidate material is present in the optimized design and used for comparing the performance of two-, three-, and four-material designs.

All simulations were implemented in MATLAB R2023b and executed on a workstation with an Intel® Xeon® w7-3545 CPU (24 cores, 2.70 GHz base) and 256 GB of RAM.

### 5.1. Bandgap maximization of adjacent eigenfrequencies

This formulation is designed to maximize the bandgap between adjacent eigenfrequencies. We have presented three numerical examples (two-, three-, and four-material design cases).

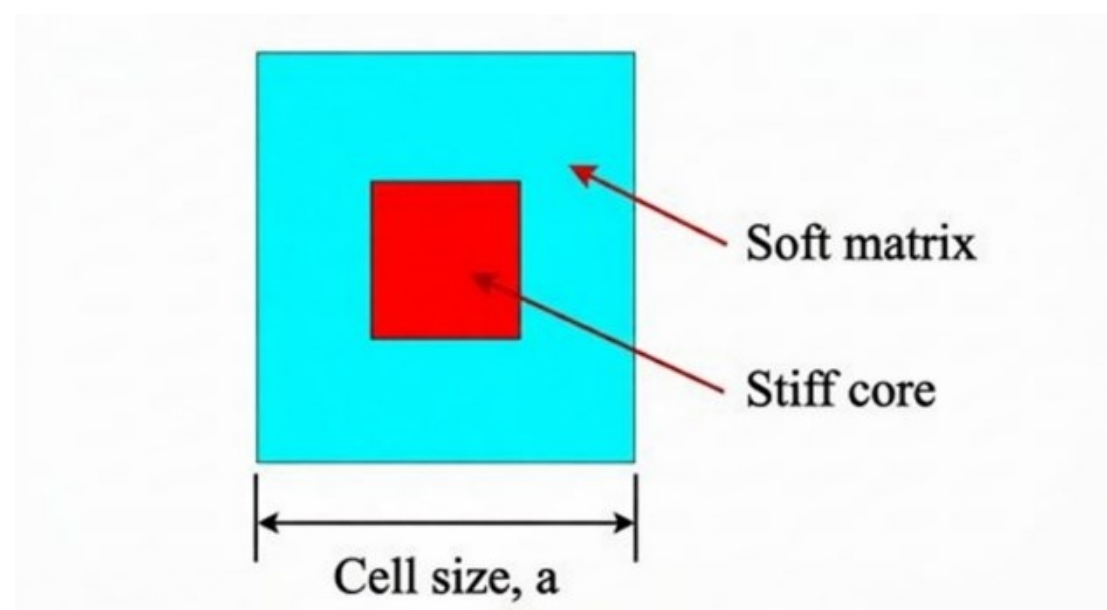

**Figure 3:** A schematic illustration of an example of a phononic unit cell consisting of two material constituents.

#### 5.1.1. Two-material phononic crystal design

In this bi-material phononic crystal, the periodic unit cell consists of epoxy (Ep) as a base matrix material and lead (Pb) as an inclusion, whose material properties are listed in ***Table 1***. The dispersion relation, as shown in ***Figure 4(a)***, is obtained by sweeping the wave vector along the boundary of the irreducible Brillouin zone, defined by the reciprocal lattice vectors [61]. The optimization problem aims to maximize the adjacent bandgap between the third and fourth eigenmodes while satisfying a 50% volume constraint for each material. The optimized design yields a maximum bandgap of $1.42\times10^4$ rad/s between these two modes, as shown in ***Figure 4(a)***. In addition to the maximum bandgap, several other bandgaps are obtained, as listed in ***Table 2***. The optimized topology corresponding to the maximum bandgap is shown in ***Figure 4(b)***, with the denser material at the center and the lighter material at the periphery.

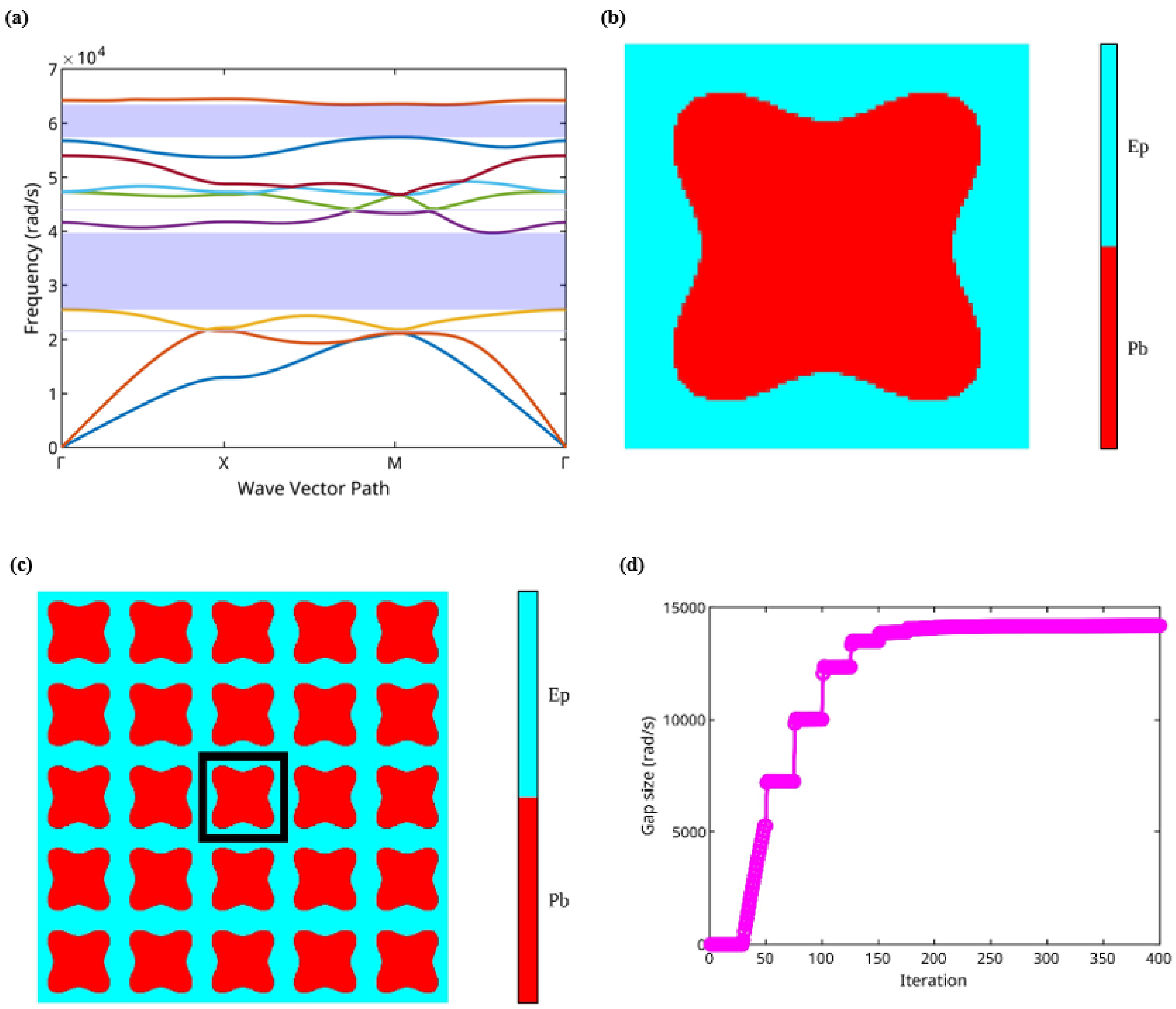

**Figure 4:** Two-material unit phononic cell for bandgap maximization between adjacent eigenfrequencies: **(a)** band structure plot, **(b)** corresponding topology of unit cell, **(c)** 5-by-5 phononic crystal, and **(d)** convergence plot

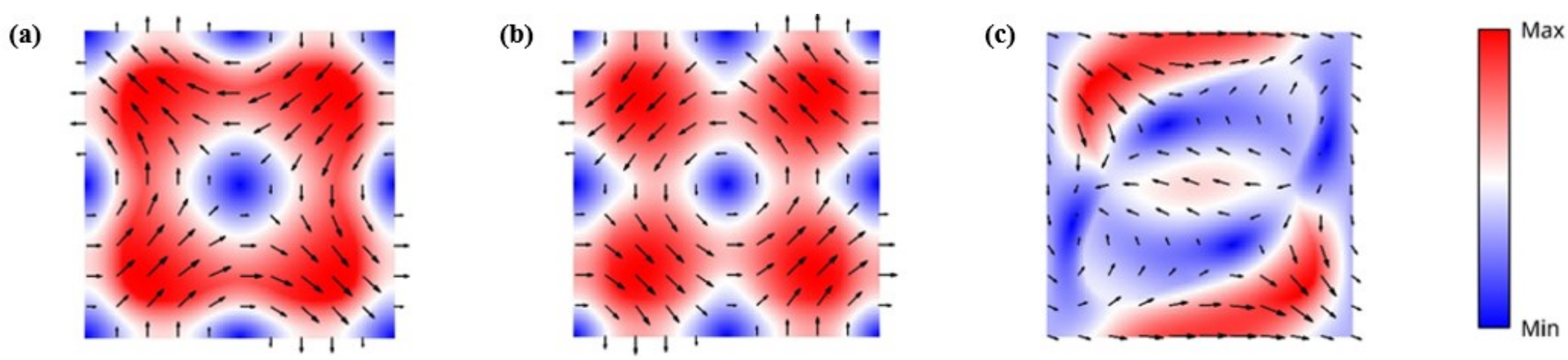

**Figure 5:** Mode shapes at the Γ point of IBZ in a two-material phononic unit cell for bandgap maximization between adjacent eigenfrequencies: **(a)** 3rd eigenmode, $2.55{\times}10^4$ rad/s, **(b)** 4th eigenmode, $4.16{\times}10^4$ rad/s, and **(c)** 5th eigenmode, $4.72{\times}10^4$ rad/s

To illustrate the periodic structure of the material, ***Figure 4(c)*** presents the full phononic crystal constructed by repeating the optimized unit cell in a 5 × 5 array, where the black square indicates a single unit cell. This 5 × 5 arrangement visualizes a phononic crystal composed of unit cells. ***Figure 4 (d)*** shows the convergence history of the optimization process. No bandgap exists during the initial iterations because not all constraints are satisfied. A bandgap emerges once the constraints are met, and then widens rapidly, after which the improvement per iteration becomes progressively smaller. Beyond 200 iterations, the bandgap width remains constant, indicating convergence to the optimum solution.

**Table 2:** Band results of optimized two-material phononic unit cell for bandgap maximization between adjacent eigenfrequencies

| Bandgap range | Bandgap width, rad/s | Bandgap starts, rad/s | Bandgap ends, rad/s |
|---|---|---|---|
| 2-3 | $1.49\times10$ | $2.18\times10^4$ | $2.18\times10^4 + 1.49\times10$ |
| 3-4 | $1.42\times10^4$ | $2.55\times10^4$ | $3.97\times10^4$ |
| 4-5 | $2.86\times10$ | $4.39\times10^4$ | $4.39\times10^4$ |
| 8-9 | $6.00\times10^3$ | $5.74\times10^4$ | $6.34\times10^4$ |

To elucidate the physical mechanisms driving bandgap attenuation, ***Figure 5*** illustrates the eigenmodes and corresponding in-plane displacement fields evaluated at the Brillouin zone center (Γ point) for the optimized two-material unit cell. The scalar color contours represent the normalized displacement magnitude, transitioning from zero-amplitude nodal regions (blue) to localized deformation peaks (red), overlaid with vector fields depicting local displacement directionality. The lower-order 3$^{rd}$ eigenmode $2.55\times10^4$ rad/s and 4$^{th}$ eigenmode $4.16\times10^4$ rad/s exhibit symmetric in-plane deformations (see ***Figures 5(a)*** and ***(b)***), characterized by identical energy distribution across all four quadrants. In contrast, the higher-order 5$^{th}$ eigenmode $4.72\times10^4$ rad/s exhibits a distinct transition toward anti-symmetric, diagonal shear motion (see ***Figure 5(c)***). This shift from lower-order symmetric expansion to higher-order anti-symmetric shear alters the local wave, showing the influence of optimized spatial redistribution of mass and stiffness on the dynamic impedance contrast required to maximize frequency separation.

### 5.1.2. Three-material phononic crystal design

For the three-material phononic crystals, the periodic unit cell consists of lead (Pb), brass (Br), and epoxy (Ep), whose material properties are listed in ***Table 1***. Here, the optimization maximizes the adjacent bandgap between the third and fourth eigenmodes, and the volume fraction of each material is 33.33%. The optimized design yields the maximum bandgap of $2.85\times10^4$ rad/s between these two modes, as shown in ***Figure 6(a)***. The third material (brass) incorporated into the unit cell has a lower density but a higher stiffness than the inclusion material (lead) used in the two-material configuration, resulting in a higher acoustic impedance. This increased impedance mismatch contributes to a wider band gap in the three-material phononic unit cell compared to the two-material configuration. It should be noted that the introduction of additional materials does not itself guarantee a wider bandgap [56]. It depends on the elastic and inertial properties of the added materials and on the volume fractions associated with them. For the numerical example considered in this study (brass as an additional material and 33.33% volume fraction for each of the three base materials), the three-material design yields a notably larger bandgap width.

In addition, ***Table 3*** shows other bandgaps. ***Figure 6 (b)*** shows the optimized topology corresponding to the maximum bandgap, with few gray elements, indicating intermediate material densities. To illustrate the periodic structure of the material, ***Figure 6(c)*** presents the full phononic crystal constructed by repeating the optimized unit cell in a 5 × 5 array, where the black square indicates a single unit cell. ***Figure 6 (d)*** shows the convergence history of the optimization process. The gap size rises rapidly during the early iterations, then exhibits several stepwise fluctuations as the design variables are updated at fixed intervals of 25 iterations. Beyond iteration 150, the gap size decreases and stabilizes, as the optimizer transitions from exploring gap-maximizing configurations toward satisfying the remaining design constraints, ultimately converging to a stable solution beyond iteration 200.

**Table 3:** Band results of optimized three-material phononic unit cell for bandgap maximization between adjacent eigenfrequencies

| Bandgap range | Bandgap width, rad/s | Bandgap starts, rad/s | Bandgap ends, rad/s |
|---|---|---|---|
| 3-4 | $2.85\times10^4$ | $3.03\times10^4$ | $5.88\times10^4$ |
| 4-5 | $7.12\times10$ | $6.36\times10^4$ | $6.37\times10^4$ |
| 7-8 | $2.01\times10^3$ | $7.75\times10^4$ | $7.95\times10^4$ |

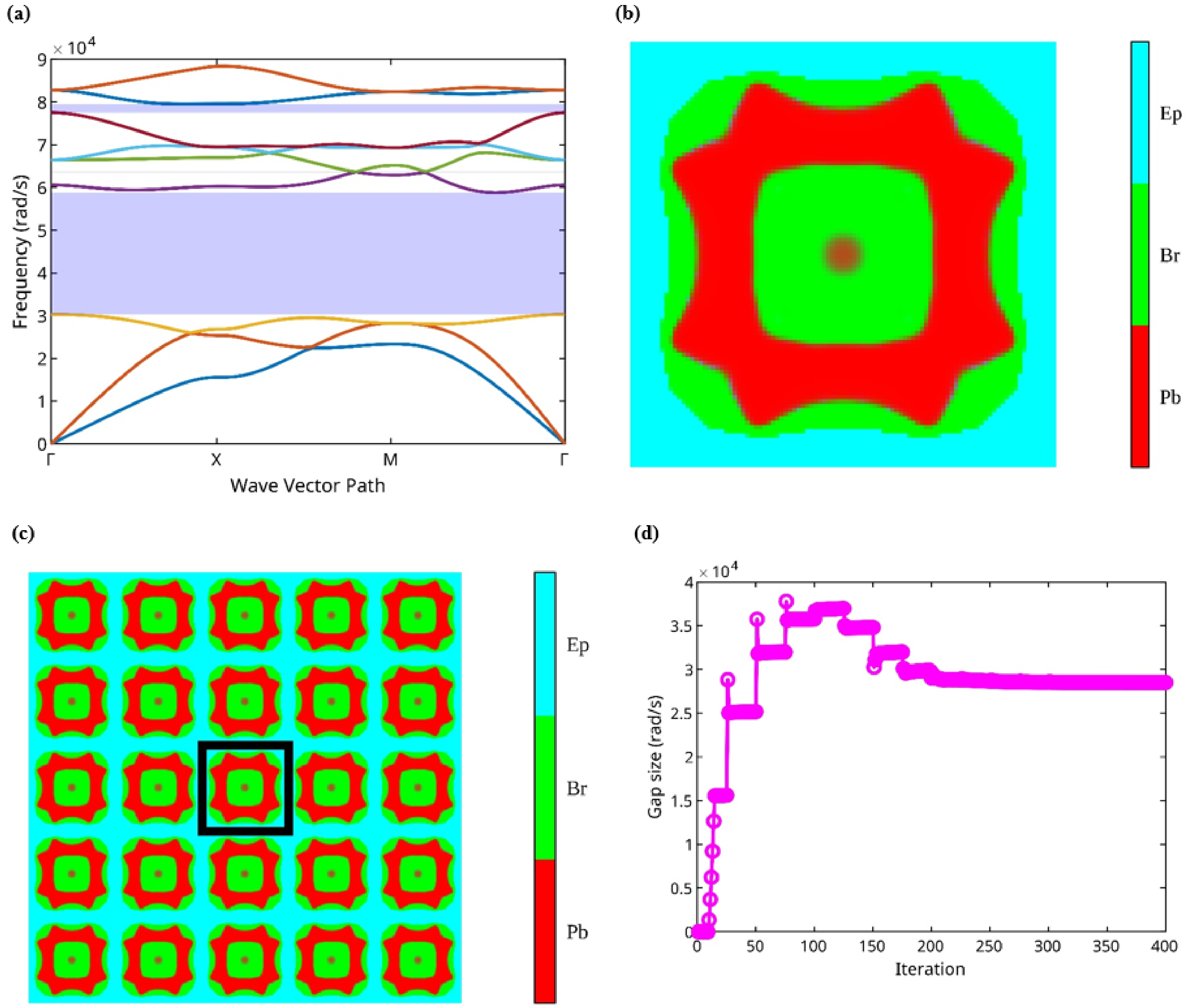

**Figure 6:** Three-material phononic unit cell for bandgap maximization between adjacent eigenfrequencies: **(a)** band structure plot, **(b)** corresponding topology of unit cell, **(c)** 5 by 5 phononic crystal, and **(d)** convergence plot

### 5.1.3. Four-material phononic crystal design

In the four-material phononic crystal design, the periodic unit cell consists of aluminum (Al), lead (Pb), brass (Br), and epoxy (Ep), whose material properties are listed in ***Table 1***. Here, the optimization problem is to maximize the adjacent bandgap between the third and fourth eigenmodes, where each material's volume fraction is fixed at 25%. The optimized design yields a maximum bandgap of $3.48 \times 10^4$ rad/s between these two modes, as shown in ***Figure 7(a)***. Incorporating a fourth material (aluminum), whose acoustic impedance is comparable to that of lead, further widens the band gap relative to the three-material configuration. The band gap is 22% wider than the three-material result and more than twice the two-material result, though the gap has also shifted upward in frequency. This indicates that even without a substantial impedance contrast between the newly added and existing materials, the increased heterogeneity within the unit cell can enhance band-gap width, largely due to the additional scattering and multi-mode coupling introduced by the extra base material.

In addition to this maximum bandgap, another bandgap exists between modes 7 and 8, as shown in ***Table 4***. ***Figure 7 (b)*** shows the optimized topology corresponding to the maximum bandgap, where some gray elements indicate intermediate material densities. ***Figure 7(c)*** presents the full phononic crystal constructed by repeating the optimized unit cell in a 5 × 5 array, where the black square indicates a single unit cell. ***Figure 7 (d)*** shows the convergence history of the optimization process. It follows the same trend as the three-material case: the gap size rises rapidly, undergoes stepwise fluctuations from design parameter updates, and finally converges to a constant bandgap width.

Table 4: Band results of optimized four-material phononic unit cell for bandgap maximization between adjacent eigenfrequencies

| Bandgap range | Bandgap width, rad/s | Bandgap starts, rad/s | Bandgap ends, rad/s |
|---|---|---|---|
| 3-4 | $3.48\times10^4$ | $3.86\times10^4$ | $7.33\times10^4$ |
| 7-8 | $1.28\times10^4$ | $9.06\times10^4$ | $1.03\times10^5$ |

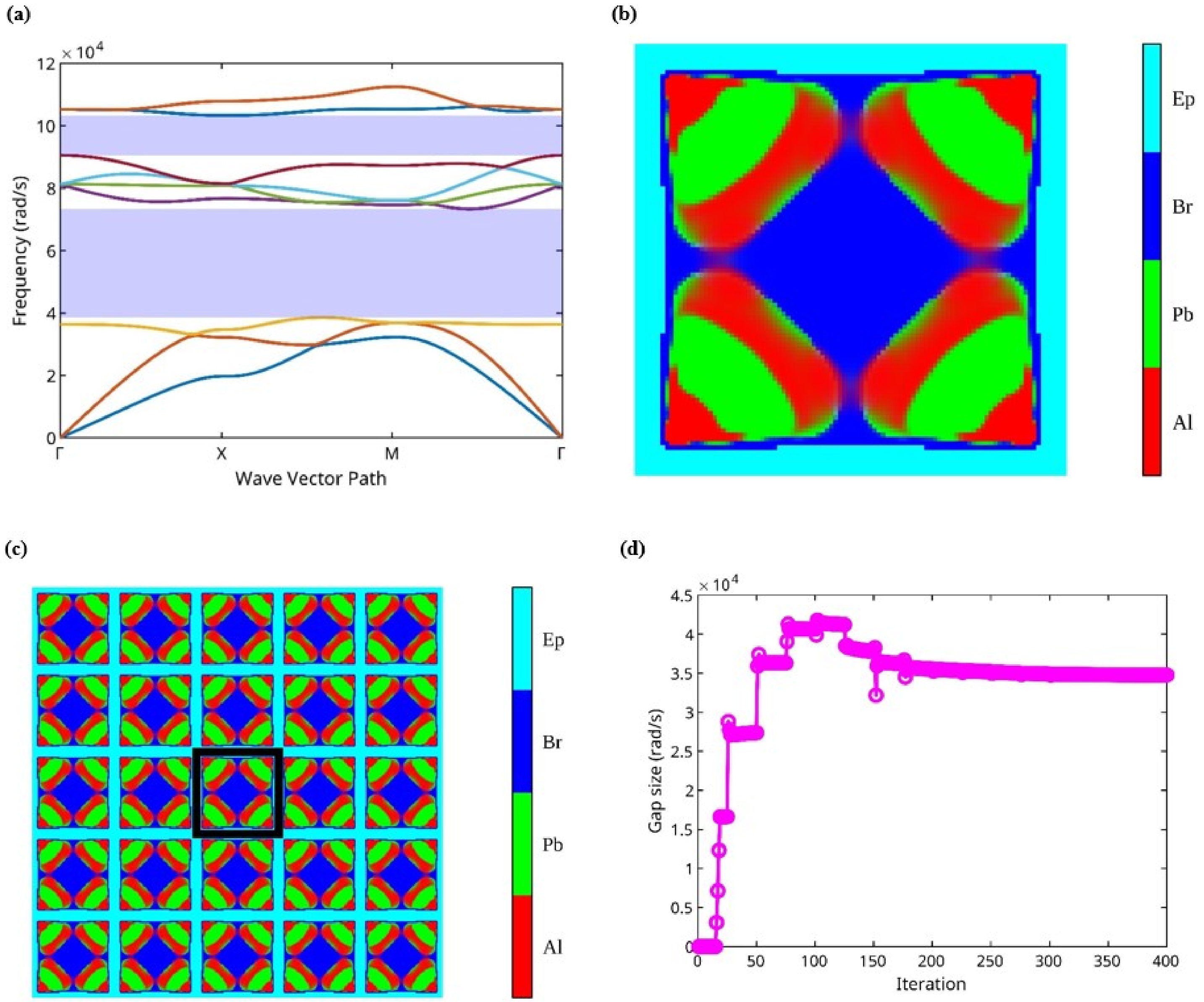

**Figure 7:** Four-material phononic unit cell for bandgap maximization between adjacent eigenfrequencies**: (a)** band structure plot, **(b)** corresponding topology of unit cell, **(c)** 5 by 5 phononic crystal, and **(d)** convergence plot

The three preceding cases demonstrate that a stiff material located at the center of the unit cell is not a prerequisite for maximizing the bandgap. Instead, the optimized material topology is strongly coupled to the volume fraction limits of the candidate material constituents.

### 5.2. Bandgap maximization of targeted frequency range

This formulation maximizes the bandgap within a prescribed frequency range and is applied to three numerical examples involving two-, three-, and four- candidate base materials.

#### 5.2.1. Two-material phononic crystal design

In this example, we apply the same material properties and volume fraction as in Section 5.1.1. We define the target bandgap range as [4.5, 5.5] kHz or [$2.83\times10^4$, $3.46\times10^4$] rad/s, with the region between the two dashed lines shown in ***Figure 8(a)***. We find that an optimized bandgap (see ***Table 5***) is wider than the target. This is because the frequency constraints are *one-sided*, requiring only that the band edges lie beyond the target bounds rather than match the bounds exactly. The unbounded width-maximizing objective drives the optimizer to keep widening the gap as long as it remains feasible under the volume constraint.

Another reason comes from the approximation error introduced by the aggregation across sampled wavevectors.

***Figures 8(a)*** and ***8(b)*** show the band structure of the optimum design and the optimized topology corresponding to the optimum bandgap, respectively. We illustrate the periodic phononic structure in ***Figure 8(c)*** with a 5 × 5 array, where the black square indicates a single unit cell. The convergence history of the optimization process is shown in ***Figure 8(d)***, in which only the active band pair, eigenmodes in red and blue colors in ***Figure 8(a)***, is tracked. During the initial 70 iterations, no bandgap is observed within the prescribed bandgap range. As optimization progresses further, the bandgap increases and becomes constant after 200 iterations.

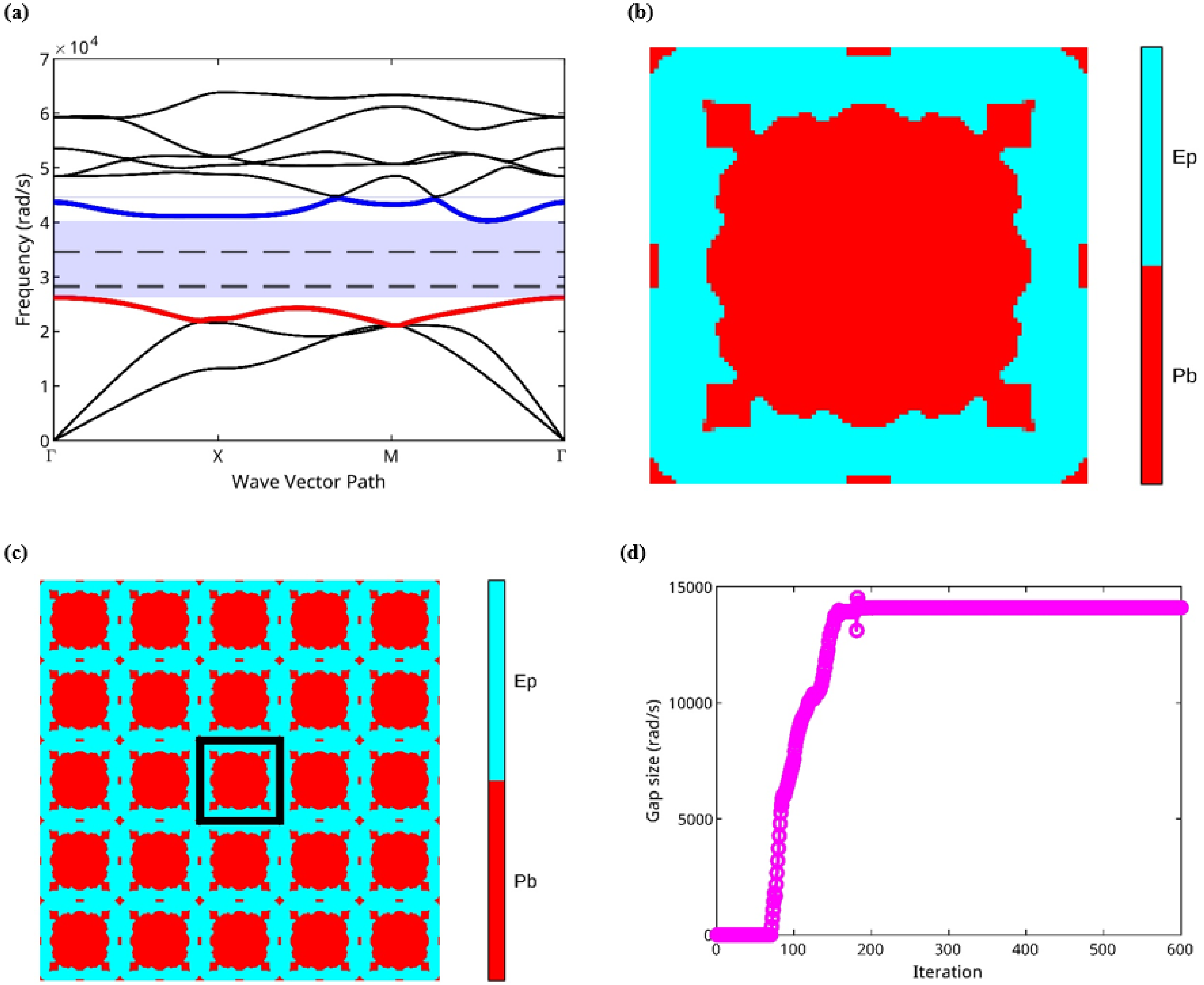

**Figure 8:** Two-material phononic unit cell for prescribed bandgap of [4.5, 5.5] kHz: **(a)** band structure plot (red and blue colors show active bandgap pair), **(b)** corresponding topology of unit cell, **(c)** 5 by 5 phononic crystal, and **(d)** convergence plot

**Table 5:** Band results of optimized two-material phononic unit cell for bandgap maximization between targeted frequency range

| Bandgap range | Bandgap width, rad/s | Bandgap starts, rad/s | Bandgap ends, rad/s |
|---|---|---|---|
| 3-4 | $1.41\times10^4$ | $2.62\times10^4$ | $4.03\times10^4$ |
| 4-5 | $1.84\times10^2$ | $4.45\times10^4$ | $4.47\times10^4$ |

### 5.2.2. Three-material phononic crystal design

To maintain consistency with the three-material example, we apply the material properties and volume-fraction constraints prescribed in Section 5.1.2 to this case. The target bandgap range is specified as [6.5, 8.5] kHz or [$4.08\times10^4$, $5.34\times10^4$] rad/s, corresponding to the region bounded by the dashed lines in ***Figure 9(a)***. As reported in ***Table 6***, the resulting bandgap exceeds the targeted bandwidth, driven by the underlying physical mechanisms detailed in ***Section 5.2.1***. ***Figures 9(a)*** and ***9(b)*** illustrate the dispersion relations and the optimized unit cell topology, respectively. Furthermore, ***Figure 9(c)*** displays the periodic phononic crystal constructed via a 5×5 array, where a single representative unit cell is highlighted by the black square.

***Figure 9 (d)*** shows the convergence history of the optimization process. The bandgap remains zero during the first 65 iterations. It significantly oscillates after the continuation parameters are updated, driving a change in material morphology that opens the bandgap. This is followed by brief oscillations as the optimizer locally readjusts to the sharpened parameters, after which the gap size decreases steadily and converges to a stable bandgap.

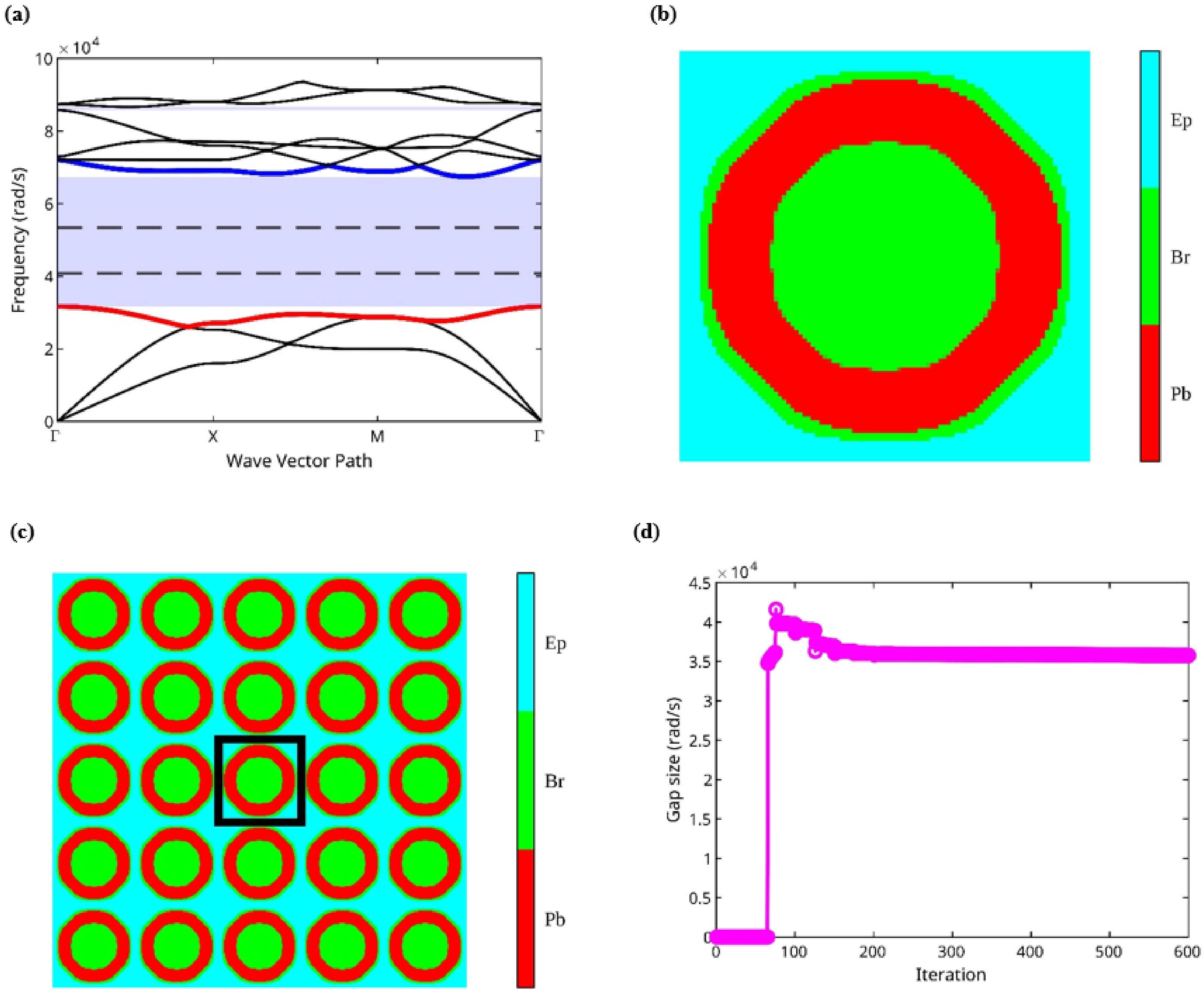

**Figure 9: Three-material phononic unit cell for prescribed bandgap of [6.5, 8.5] kHz: (a)** band structure plot (red and blue colors show active bandgap pair), **(b)** corresponding topology of unit cell, **(c)** 5 by 5 phononic crystal, **(d)** convergence plot

**Table 6:** Band results of optimized three-material phononic unit cell for bandgap maximization between targeted frequency range

| Bandgap range | Bandgap width, rad/s | Bandgap starts, rad/s | Bandgap ends, rad/s |
|---|---|---|---|
| 3-4 | $3.58\times10^4$ | $3.17\times10^4$ | $6.74\times10^4$ |
| 7-8 | $8.80\times10^2$ | $8.58\times10^4$ | $8.66\times10^4$ |

### 5.2.3. Four-material phononic crystal design

For consistency, we retain the material properties and volume fraction specified in Section 5.1.3. The target bandgap spans [7, 9] kHz or [$4.40\times10^4$, $5.65\times10^4$] rad/s, indicated by the dashed lines in ***Figure 10(a)***. The optimized design achieves a wider bandgap than the target (see ***Table** 7*). ***Figures 10(a)*** and ***10(b)*** show the dispersion diagram and the optimized unit cell topology. Here, the alternating acoustic impedance across the successive radial layers opens and widens the targeted bandgap. ***Figure 10(c)*** shows the corresponding periodic phononic crystal assembled as a 5×5 array, with one representative unit cell outlined in black. ***Figure 10(d)*** presents the convergence history. No bandgap opens within the target range during the first 78 iterations. At the 79th iteration, the optimization first satisfies all constraints, producing an abrupt increase in bandgap width. The bandgap proceeds to fluctuate as the design converges to a local optimum. After the 200th iteration, the generated bandgap remains stable.

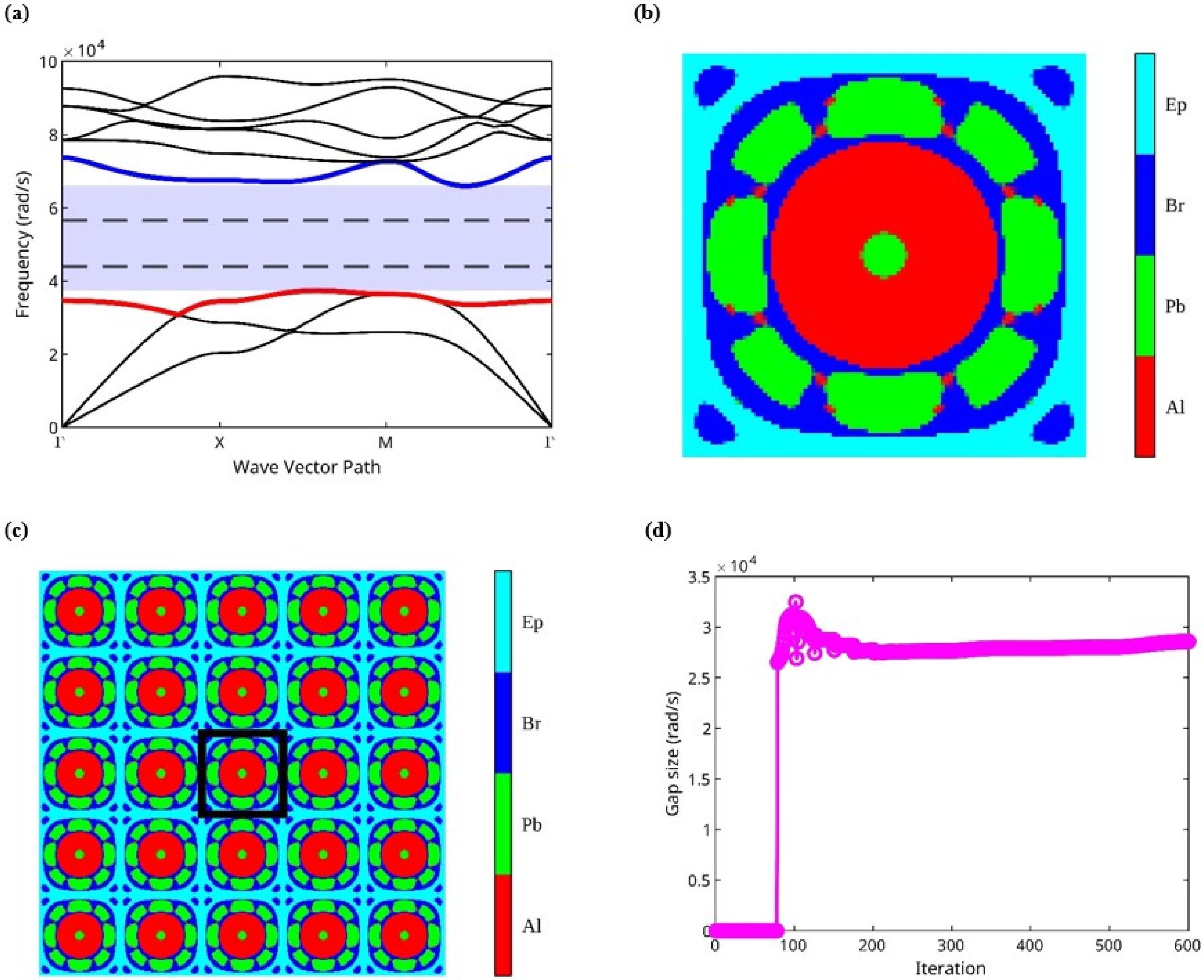

**Figure 10:** Four-material phononic unit cell for prescribed bandgap of [7, 9] kHz: **(a)** band structure plot (red and blue colors show active bandgap pair), **(b)** corresponding topology of unit cell, **(c)** 5 by 5 phononic crystal, and **(d)** convergence plot

**Table 7:** Band results of optimized four-material phononic unit cell for bandgap maximization between targeted frequency range

| Bandgap range | Bandgap width, rad/s | Bandgap starts, rad/s | Bandgap ends, rad/s |
|---|---|---|---|
| 3-4 | $2.86\times10^4$ | $3.74\times10^4$ | $6.60\times10^4$ |

## 6. Numerical Validation

To verify our designs, transmission loss in a 5×5 phononic crystal is calculated in the frequency domain by using the solid mechanics module in COMSOL Multiphysics. Unlike the dispersion analysis, which assumes an infinite periodic medium, the finite periodic array adopted here allows the attenuation performance of the phononic crystal to be evaluated under realistic boundary conditions. We attach perfectly matched layers (PMLs) with the same material properties as the host medium adjacent to the input and output regions. These layers minimize boundary reflections, thereby preventing them from contaminating the transmitted response. The transmission loss spectrum serves as a direct and quantitative indicator of the frequency ranges over which elastic wave propagation is suppressed. The transmission loss [62, 63] can be defined as ***Equation (46)***:

$$\mathrm{TL} = 20 * \log_{10}\left(\frac{\mathrm{d_{out}}}{\mathrm{d_{in}}}\right), \tag{46}$$

where $d_{in}$ is the prescribed displacement imposed at the input boundary of the phononic crystal array, applied along the x-direction only with a magnitude of 1 mm, and $d_{out}$ is the magnitude of the resultant in-plane displacement evaluated at the output boundary of the array, obtained from its x- and y-components. The transmission loss in decibels (dB) is calculated, where high transmission loss values correspond to elastic wave propagation and low transmission loss values correspond to blocked waves.

### 6.1. Transmission loss for bandgap maximization between adjacent eigenfrequencies

Three numerical examples corresponding to two-, three-, and four-material design cases are investigated to validate the proposed topology optimization for bandgap maximization between adjacent eigenfrequencies. The transmission-loss performance of the optimized configurations is evaluated in the frequency domain using COMSOL Multiphysics.

#### 6.1.1. Two-material phononic crystal

The optimum bandgap obtained between bands 3 and 4 of the two-material phononic crystal is $1.42\times10^4$ rad/s ($2.55\times10^4$ rad/s to $3.97\times10^4$ rad/s). COMSOL also shows that, within this range, the wave cannot propagate (see ***Figure 11(a)***), which is also confirmed by the transmission loss obtained from the frequency analysis. The input displacement is not perfectly constant at 1 mm across the frequency range, as the response is recorded at an edge on the input face of the finite periodic structure rather than at the excitation source itself. At this edge, the total field is the superposition of the incident wave and the wave reflected from the periodic assembly. The plot shows that there was zero displacement in the bandgap region, and outside the bandgap region, there is non-zero displacement. Similarly, transmission loss shows negative transmission loss in the bandgap regions, as shown in the red square region in ***Figure 11(b)***.

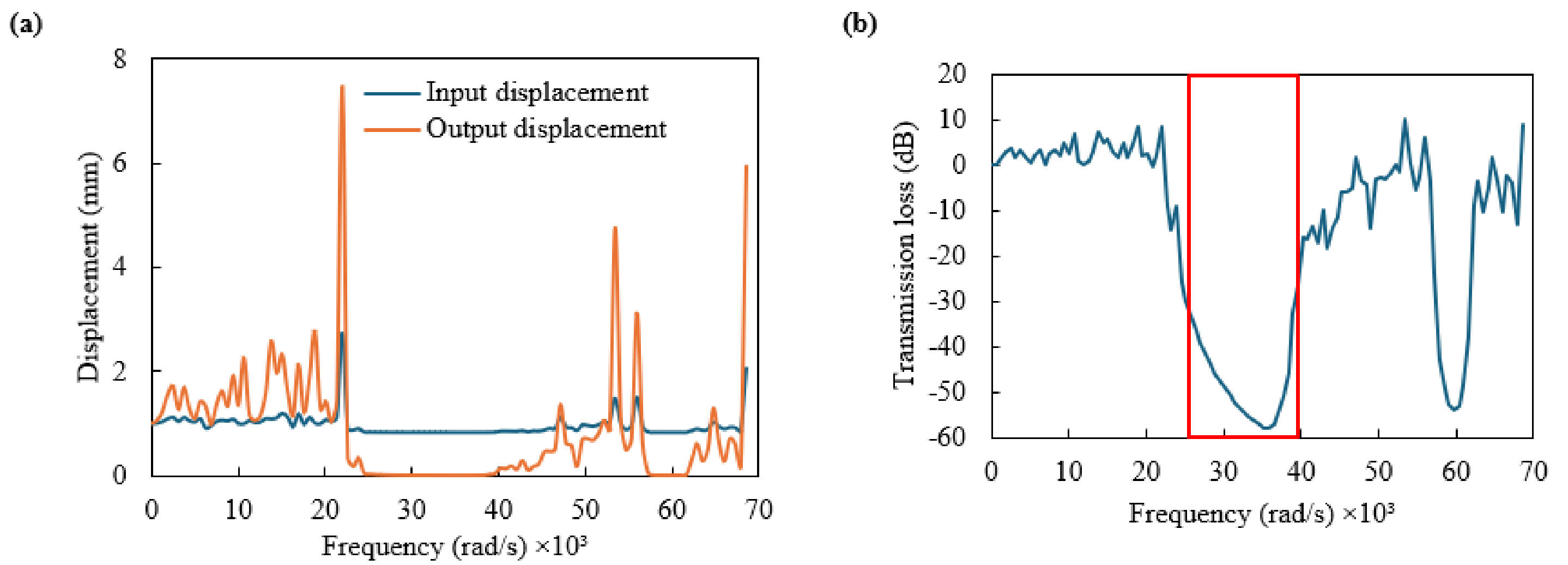

**Figure 11:** Transmission loss analysis of a two-material phononic crystal for bandgap maximization between adjacent eigenfrequencies: **(a)** input vs output displacement field, and **(b)**Transmission loss

The steady-state displacement fields in ***Figure 12*** confirm the bandgap predicted by the dispersion analysis. At 5001 Hz ($3.14\times10^4$ rad/s), inside the gap, waves scattered at successive layers of inclusions interfere destructively, and the response decays (see ***Figure 12(a)***), remaining confined to the first two unit cells as the amplitude falls from $1.0\times10^{-3}$ m at the driven edge to below $1.0\times10^{-4}$ m by the third column. This decay is reactive rather than dissipative, as our model is undamped. At 1501 Hz ($9.43\times10^3$ rad/s), outside the gap, waves propagate, and energy reaches the far boundary on the right side (see ***Figure 12(b)***), where the alternating displacement fields indicate a stable wave propagation pattern.

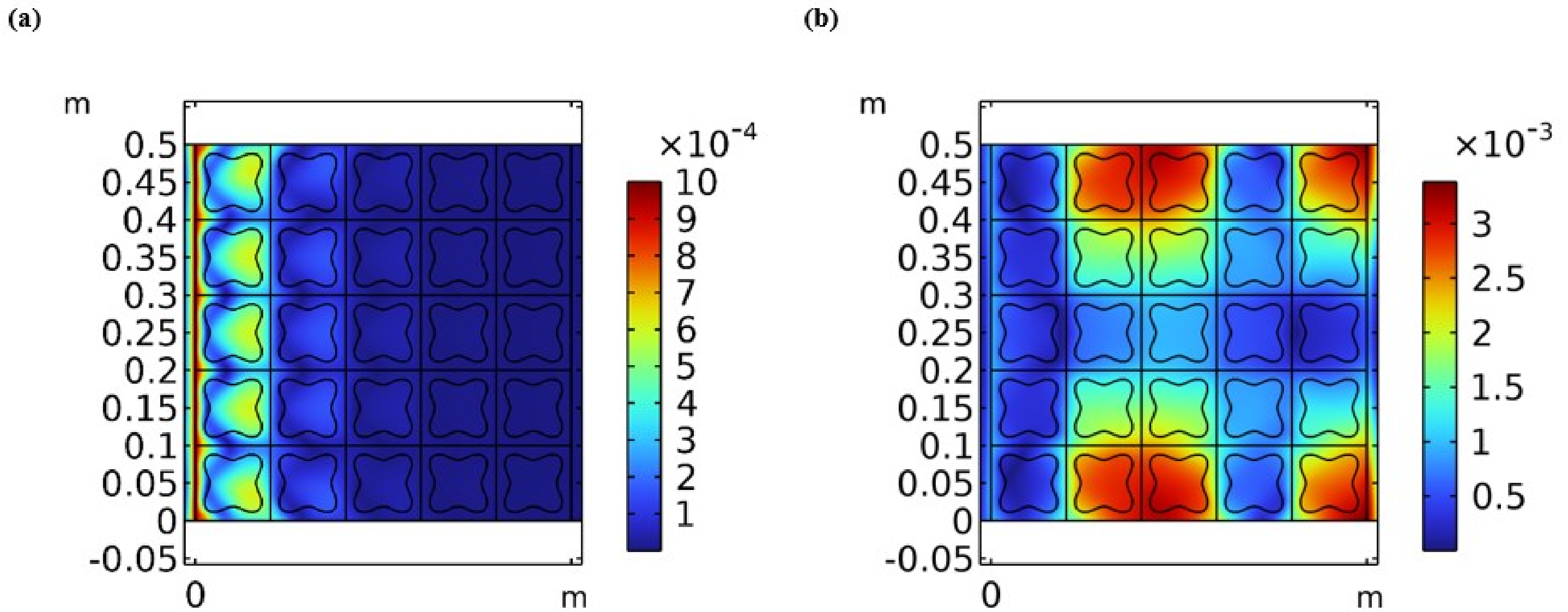

**Figure 12:** Displacement field in a two-material phononic crystal for bandgap maximization between adjacent eigenfrequencies: **(a)** at 5001 Hz (inside bandgap), and **(b)** at 1501 Hz (outside bandgap), where 'm' indicates the length unit 'meter' and the color encodes displacement magnitude.

### 6.1.2. Three-material phononic crystal

The optimum bandgap obtained between bands 3 and 4 of the three-material phononic crystal is $2.85\times10^4$ rad/s ($3.03\times10^4$ rad/s to $5.88\times10^4$ rad/s). The dispersion result is verified in COMSOL through frequency-domain analysis, where a 1 mm excitation is applied at the left boundary, and the transmitted displacement is monitored at the opposite edge. Within the optimized bandgap, the transmitted amplitude decreases by several orders of magnitude and is close to zero on the plot, whereas a significant transmission is observed outside the gap (***Figure 13(a)***). The corresponding transmission-loss curve in ***Figure 13(b)*** exhibits a pronounced attenuation band whose limits closely match the optimized bandgap, marked by the red box. The attenuation band is wider than that of the two-material crystal. The consistency between the Floquet-based unit-cell analysis and the finite-crystal forced response confirms our designed phononic bandgap.

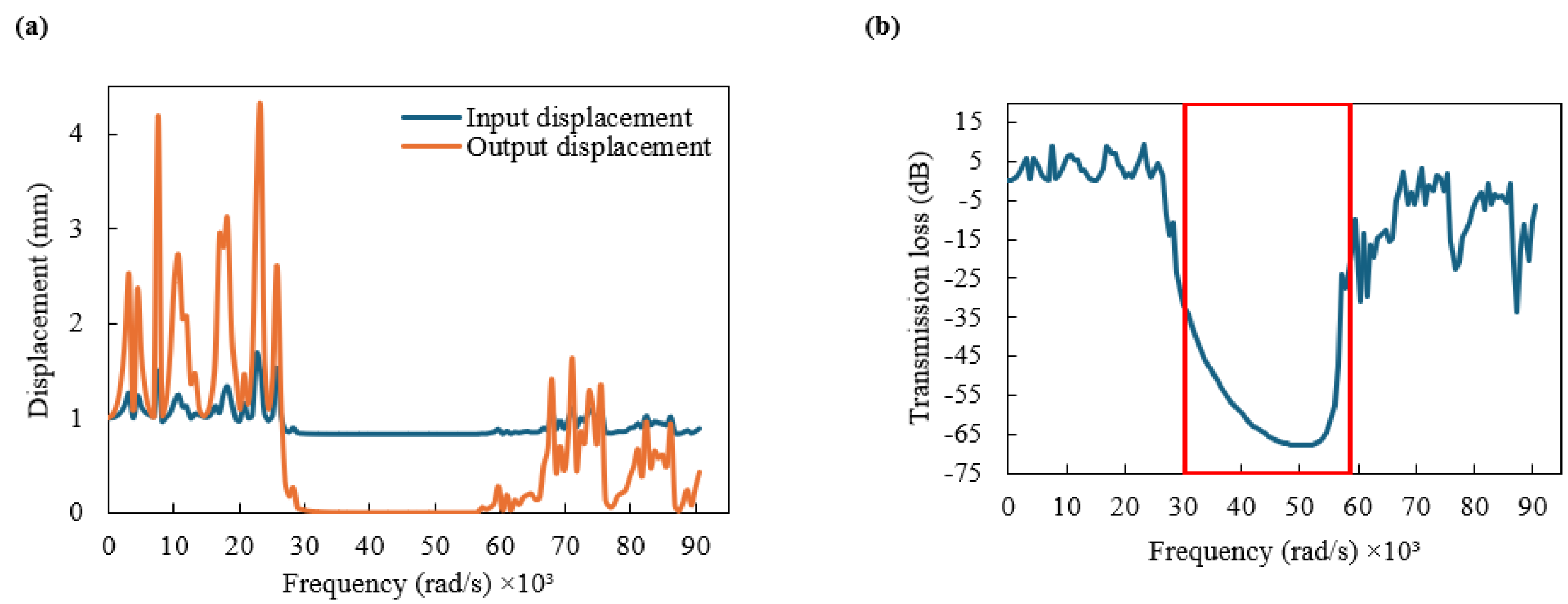

**Figure 13:** Transmission loss analysis of three-material phononic crystal for bandgap maximization between adjacent eigenfrequencies: **(a)** input vs output displacement field, and **(b)**Transmission loss

The steady-state displacement fields in the three-material phononic crystal show behavior similar to that described in Section 6.1.1. At 7001 Hz ($4.40\times10^4$ rad/s), inside the gap, the wave does not propagate, and the field decays within the first few unit cells as seen in ***Figure 14(a)***. At 3001 Hz ($1.90\times10^4$ rad/s), outside the gap, the wave propagates through the full domain and reaches the far boundary (see ***Figure 14(b)***).

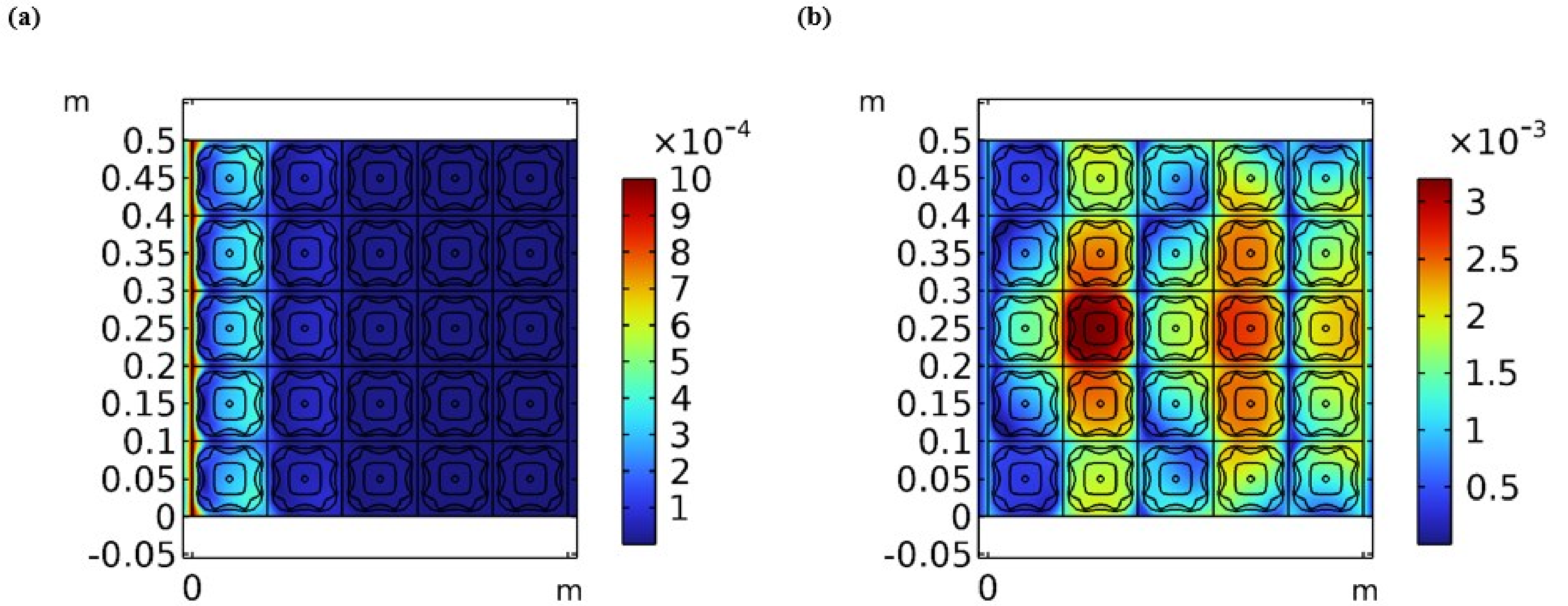

**Figure 14:** Displacement field in a three-material phononic crystal for bandgap maximization between adjacent eigenfrequencies: **(a)** at 7001 Hz (inside bandgap), and **(b)** at 3001 Hz (outside bandgap), where 'm' indicates the length unit 'meter,' and the color encodes displacement magnitude.

### 6.1.3. Four-material phononic crystal

The optimum bandgap obtained between bands 3 and 4 of the four-material phononic crystal is $3.48\times10^4$ rad/s ($3.86\times10^4$ rad/s to $7.34\times10^4$ rad/s). COMSOL shows that, within this range, the wave cannot propagate, as shown in ***Figure 15(a)***, which is confirmed by the transmission loss obtained from the frequency analysis. An initial excitation of 1 mm is given at the left boundary of the phononic crystal, and the output plot shows that there is zero displacement in the bandgap region, and outside the bandgap region, there is non-zero displacement. Similarly, transmission loss shows negative sound pressure in the bandgap regions. Here, the transmission loss region is even wider than those of the two- and three-material phononic crystals. This is the widest attenuation zone among all configurations considered in the present study, indicating that adding a fourth constituent material provides greater freedom to tune the impedance contrast within the unit cell. The red square region in ***Figure 15(b)*** is the bandgap region obtained after optimization.

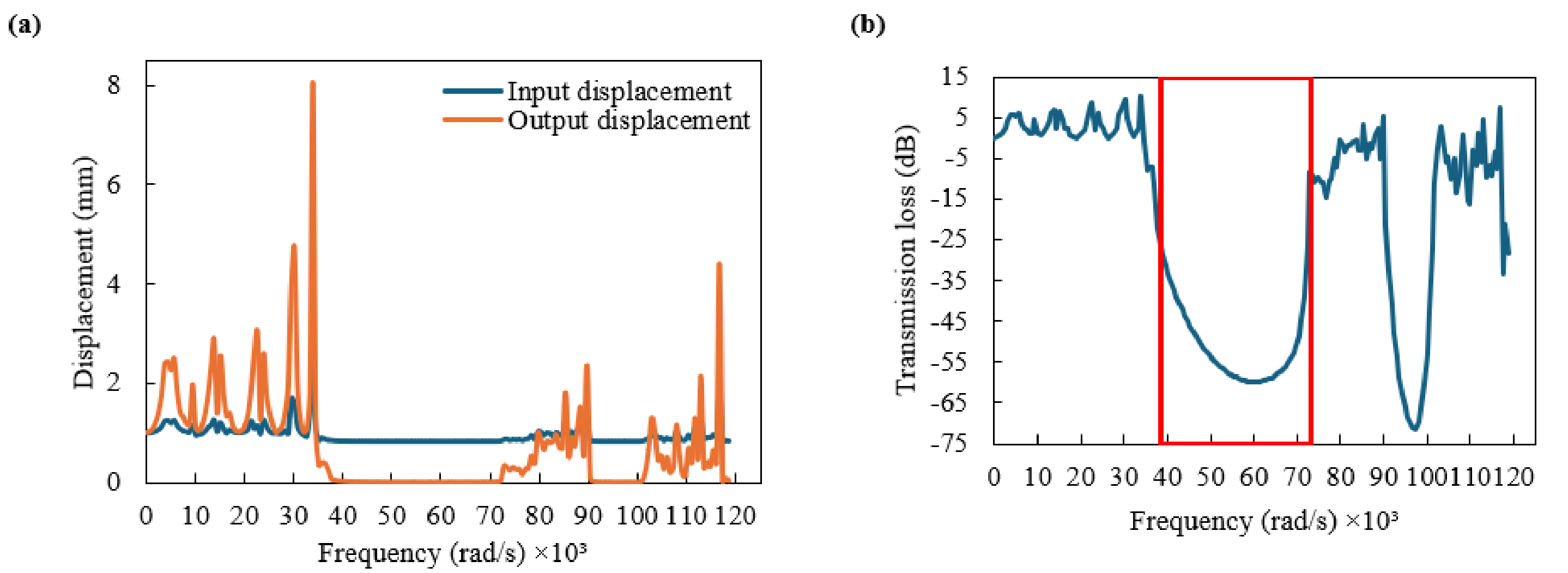

**Figure 15:** Transmission loss analysis of four-material phononic crystal for bandgap maximization between adjacent eigenfrequencies: **(a)** input vs output displacement field, and **(b)**Transmission loss

***Figure 16*** further illustrates the wave-transmission behavior at two representative frequencies. At 7501 Hz ($4.71\times10^4$ rad/s), the field is confined to the first few unit cells and does not reach the far boundary, consistent with operation within the bandgap in ***Figure 16(a)***. In contrast, at 4001 Hz ($2.51\times10^4$ rad/s), the excitation propagates across the entire domain in ***Figure 16(b)***. Together with the results in Sections 6.1.1 and 6.1.2, these observations confirm that the 5×5 array effectively suppresses transmission within the targeted bandgaps for all material configurations considered.

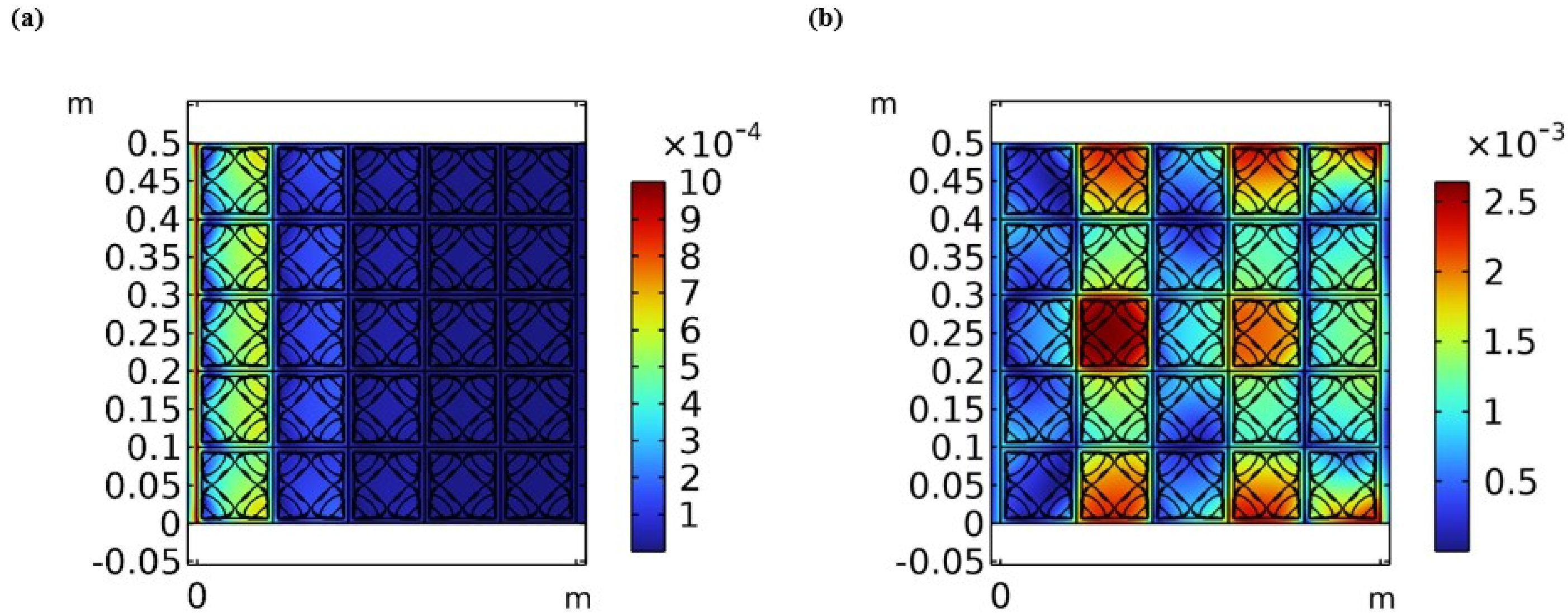

**Figure 16:** Displacement field in four-material phononic crystal for bandgap maximization between adjacent eigenfrequencies: **(a)** at 7501 Hz (inside bandgap), **(b)** at 4001 Hz (outside bandgap), where 'm' indicates the length unit 'meter,' and the color encodes displacement magnitude.

### 6.2. Transmission loss for bandgap maximization between target frequencies

Three numerical examples corresponding to two-, three-, and four-material designs are validated through the transmission-loss spectrum obtained from COMSOL Multiphysics. To verify the optimization, we perform a frequency-domain analysis under boundary conditions and initial excitation similar to those specified in the first formulation.

#### 6.2.1. Two-material phononic crystal

For the two-material case, the optimized bandgap satisfies the prescribed frequency range. Validation using COMSOL Multiphysics confirms that wave propagation is effectively suppressed within this range, as seen in ***Figure 17(a)***, further supported by the transmission-loss results from the frequency analysis. The red square region in ***Figure 17(b)*** represents the optimized bandgap.

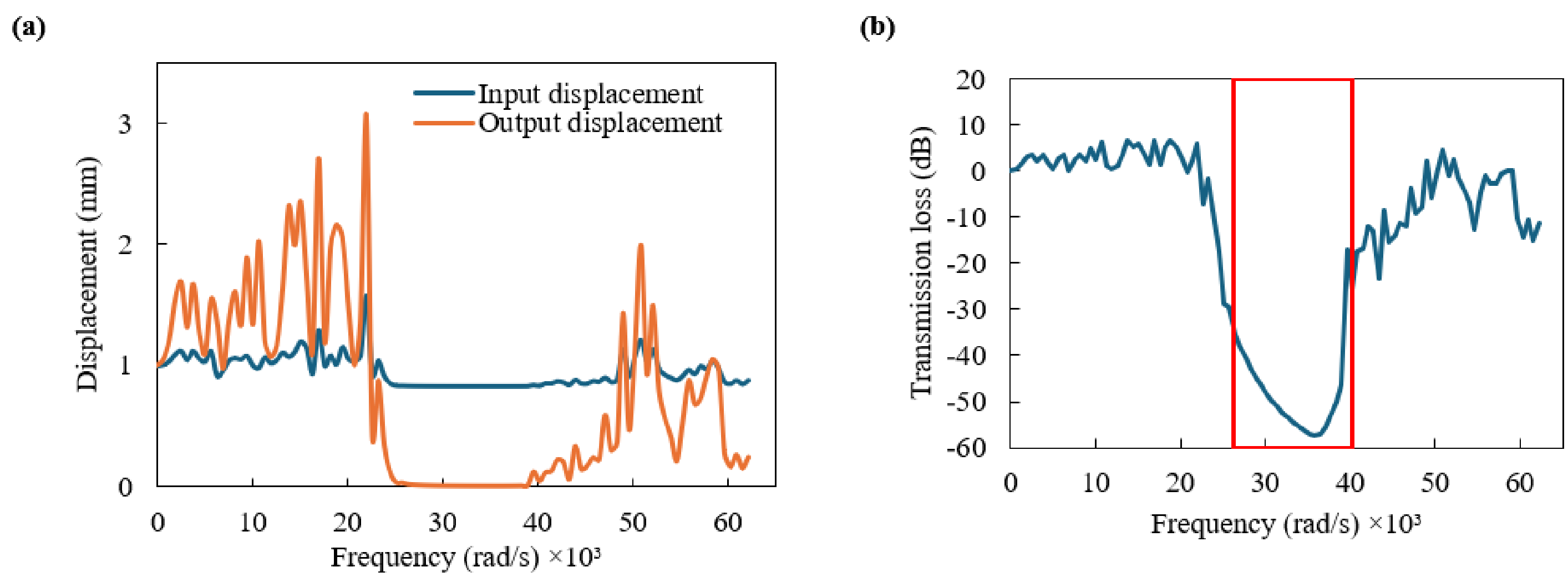

**Figure 17:** Transmission loss analysis of a two-material phononic crystal for bandgap maximization between target frequencies: **(a)** input vs output displacement field **(b)**Transmission loss

The steady-state displacement field confirms the optimized bandgap in the physical domain. For a given input applied at the left boundary, the response at 5001 Hz ($3.14\times10^4$ rad/s), which lies inside the bandgap, shows that the wave cannot propagate through the phononic crystal (see ***Figure 18(a)***); the amplitude decays within the first few unit cells, as expected of a Bragg-type stop band. At 1501 Hz ($9.43\times10^3$ rad/s), which lies outside the bandgap, the wave propagates through the entire array without significant attenuation (see ***Figure 18(b)***). This confirms that the optimization has maximized the gap within the specified target frequency range without affecting the pass bands outside it.

#### 6.2.2. Three-material phononic crystal

From this formulation, the optimum bandgap satisfies the prescribed bandgap range, and COMSOL also shows that, within this range, waves cannot propagate, as in ***Figure 19(a)***, which is supported by the transmission loss obtained from frequency analysis. The red square region in ***Figure 19(b)*** is the bandgap region obtained after optimization.

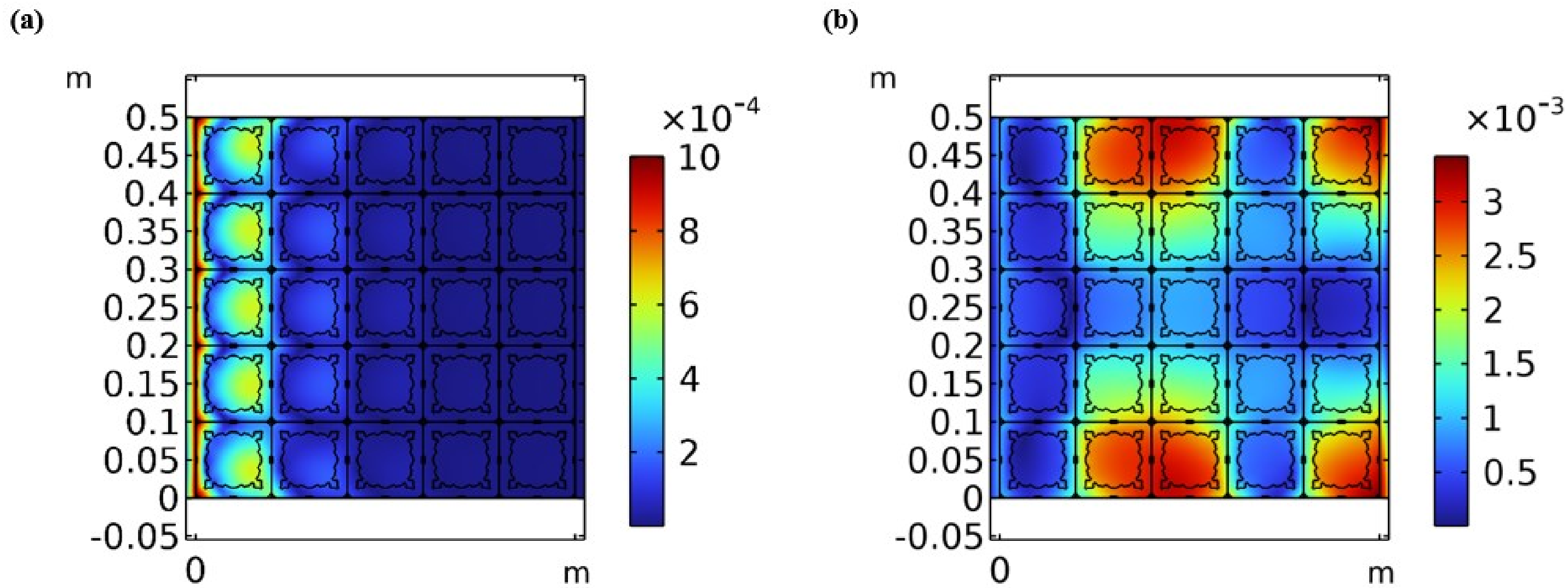

**Figure 18:** Displacement field in a two-material phononic crystal for bandgap maximization between target frequencies: (**a**) at 5001 Hz (inside bandgap), **(b)** at 1501 Hz (outside bandgap), where 'm' indicates the length unit 'meter,' and the color encodes displacement magnitude.

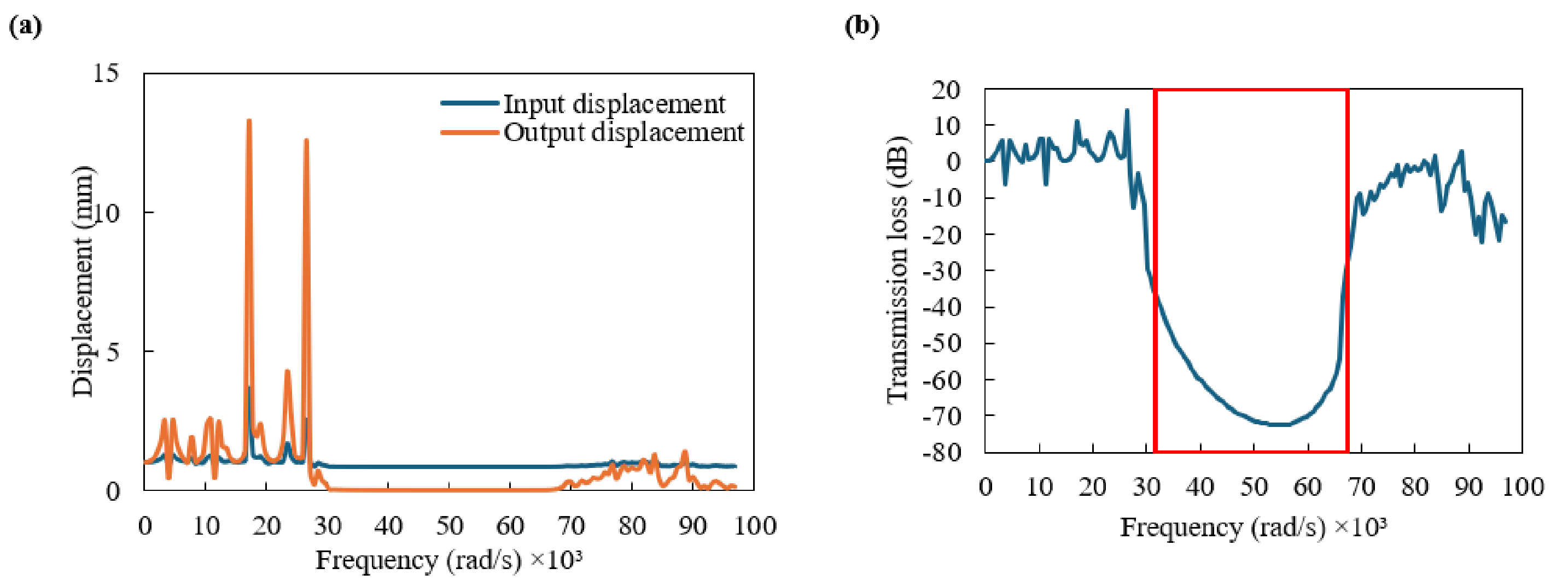

**Figure 19:** Transmission loss analysis of three-material phononic crystal for bandgap maximization between target frequencies: **(a)** input vs output displacement field **(b)**Transmission loss

The steady-state displacement field shows that, for a given input, at 7001 Hz ($4.40\times10^4$ rad/s), which lies within the prescribed range [6.5, 8.5] kHz, the wave cannot propagate through the phononic crystal (see ***Figure 20(a)***), and at 3101 Hz ($1.95\times10^4$ rad/s), which is outside the bandgap region, the wave propagates in ***Figure 20(b)***.

### 6.2.3. Four-material phononic crystal

The optimized bandgap obtained from this formulation satisfies the prescribed frequency range [7, 9] kHz. Validation using COMSOL Multiphysics confirms that wave propagation is effectively suppressed within this range in ***Figure 21(a)***, as further confirmed by the transmission-loss spectrum from frequency-domain analysis. The red box in ***Figure 21(b)*** indicates the bandgap obtained from the optimization.

The steady-state displacement field indicates that, for a given input displacement, wave propagation is inhibited at 8001 Hz ($5.03\times10^4$ rad/s), which lies within the prescribed range [7, 9] kHz in ***Figure 22(a)***. In contrast, at 4001 Hz ($2.51\times10^4$ rad/s), corresponding to a frequency outside the bandgap, the wave propagates through the phononic crystal as shown in ***Figure 22(b)***.

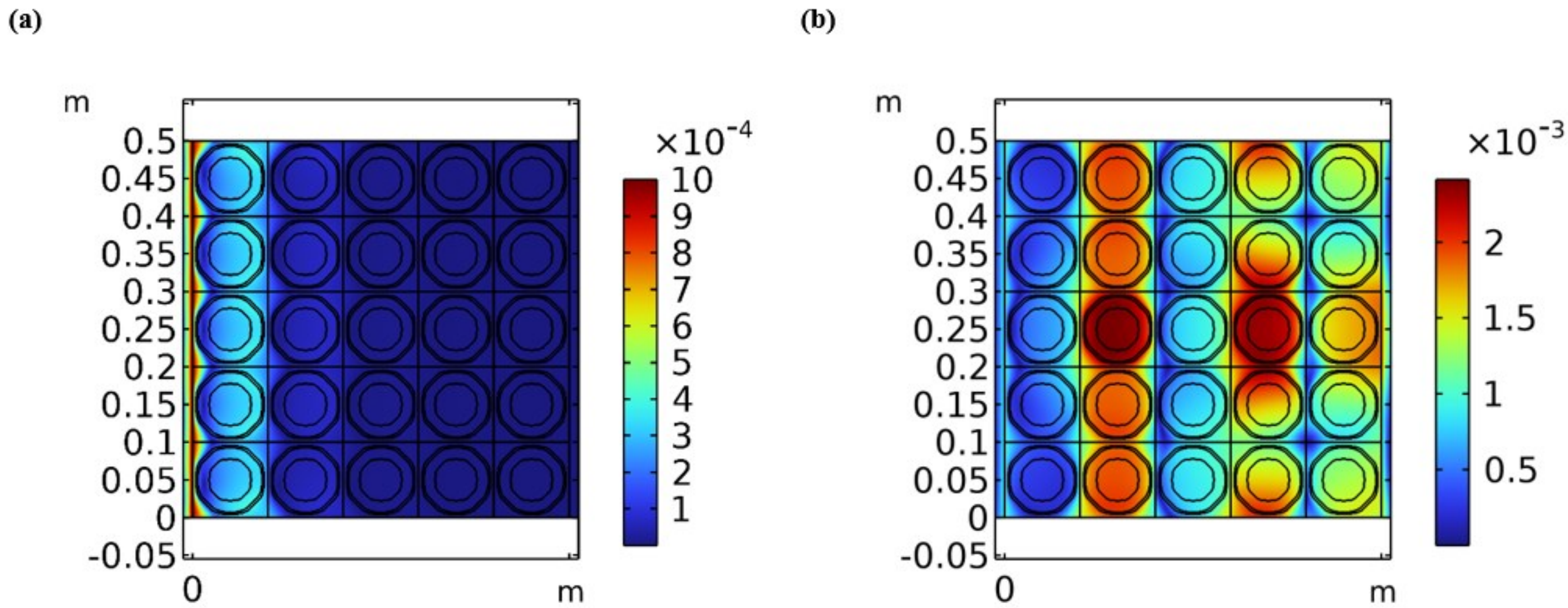

**Figure 20:** Displacement field in three-material phononic crystal for bandgap maximization between target frequencies: (**a**) at 7001 Hz (inside bandgap), and **(b)** at 3101 Hz (outside bandgap), where 'm' indicates the length unit 'meter,' and the color encodes displacement magnitude.

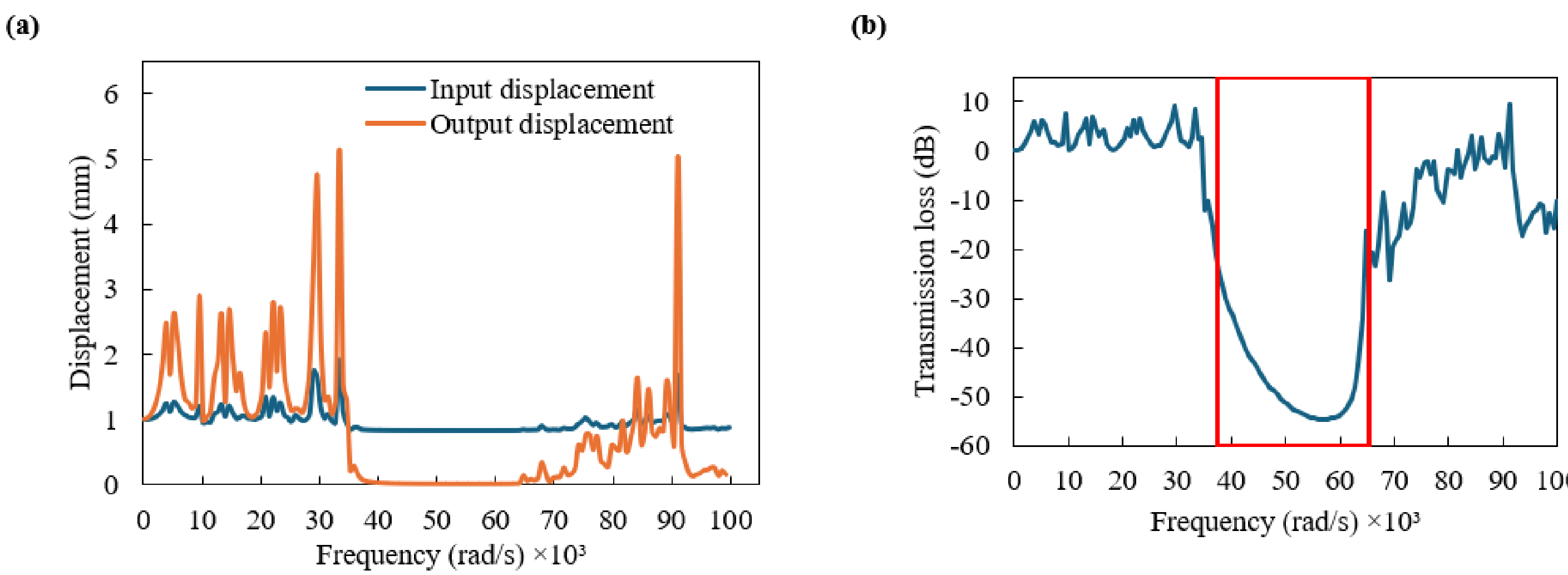

**Figure 21:** Transmission loss analysis of four-material phononic crystal for bandgap maximization between target frequencies: **(a)** input vs output displacement field **(b)**Transmission loss

## 7. Conclusions

In this paper, we propose a multivariate weight-based topology optimization method for phononic crystals to maximize bandgaps, where we formulate a mapping to project the design variables of an arbitrary number of candidate base materials to the field of effective density and an effective stiffness tensor. Specifically, we develop two formulations for the maximum bandgap in multi-material phononic crystals. The first formulation maximizes the bandgap between two adjacent eigenmodes. The second formulation tailors the bandgap to a user-defined target frequency range via a dynamic band-pairing scheme, in which the pair of adjacent bands is adaptively identified during optimization until the resulting bandgap fully encloses the targeted frequency range.

We applied both formulations to periodic multi-material phononic crystals. To investigate the influence of the number of constituent base materials on bandgaps, we considered three configurations: bi-material, tri-material, and quad-material designs, with an equal volume fraction imposed on each material. In all cases, the optimized designs satisfy the prescribed volume constraints while significantly enhancing the targeted bandgap, and the prescribed-bandgap formulation successfully places the gap over the targeted frequency range. We validate the optimized phononic crystals in COMSOL through transmission-loss analysis, confirming our proposed design formulations.

Several further extensions remain open. One such direction is the simultaneous control of multiple bandgaps, in which gaps between several adjacent band pairs are opened within a single optimization rather than one targeted pair at a time. Another is generalizing both formulations to three-dimensional unit cells and to fully coupled in-plane and out-of-plane wave propagation. Finally, incorporating manufacturing constraints into the optimization and experimentally validating the resulting designs would strengthen the framework's practical relevance.

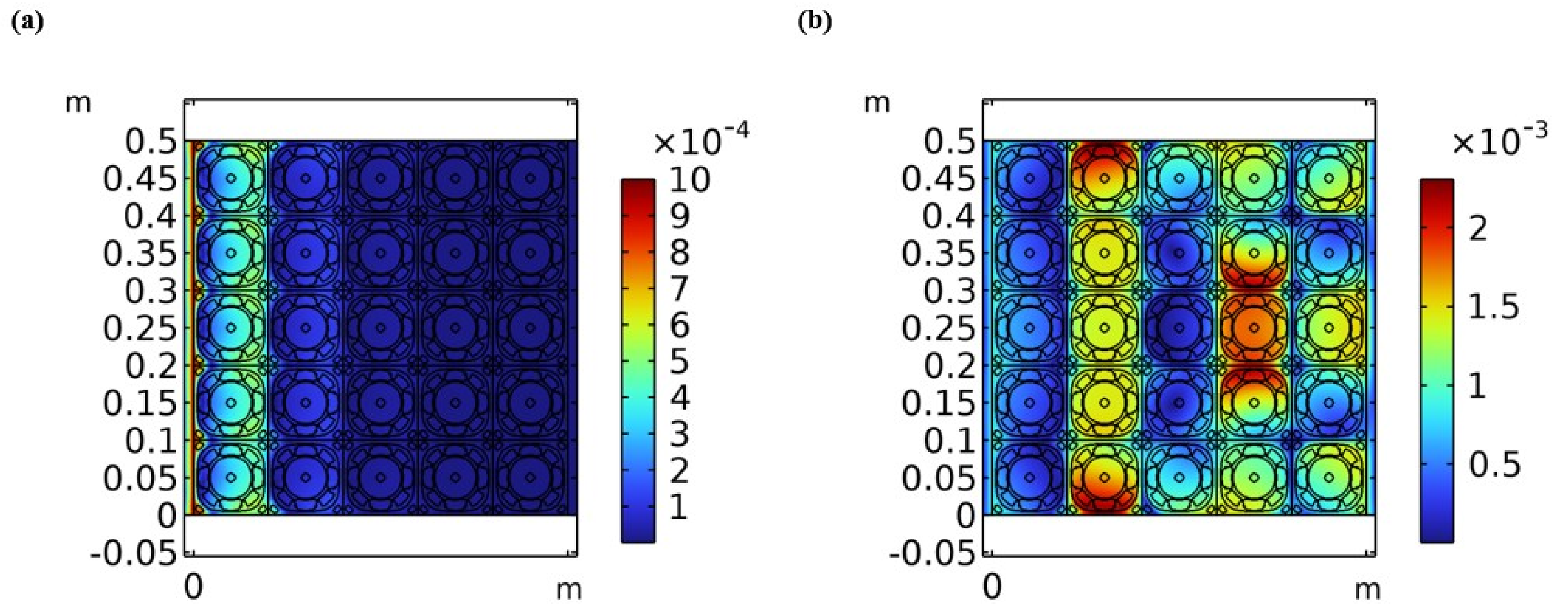

**Figure 22:** Displacement field in four-material phononic crystal for bandgap maximization between target frequencies: **(a)** at 8001 Hz (inside bandgap), and **(b)** at 4001 Hz (outside bandgap), where 'm' indicates the length unit 'meter,' and the color encodes displacement magnitude

## ACKNOWLEDGEMENTS

The authors would like to thank the constructive suggestions from anonymous reviewers. The authors would like to thank Prof. Krister Svanberg for the MATLAB code for the MMA optimizer. Shiguang Deng would like to thank the support of the Faculty Startup Grant, the New Faculty Research Development Grant, and the General Research Grant from the University of Kansas. The Los Alamos unlimited release number is LA-UR-26-28858.

## APPENDIX A: Sensitivity Derivation for Bandgap

### 1. Gradient of weights

i. For material 1, gradients of unnormalized weights in tensor form can be computed as:

$$\frac{\partial \mathbf{w}_1^{un}}{\partial \boldsymbol{\phi}_1} = r\boldsymbol{\phi}_1^{r-1}. \tag{A1}$$

ii. For material $2 \leq m \leq n_{Mat} - 1$, gradients of unnormalized weights are given in tensor form:

a. When the design field (f) is less than the number of materials (m):

$$\frac{\partial \mathbf{w}_m^{un}}{\partial \boldsymbol{\phi}_f} = \left(\prod_{\substack{j=1 \\ j\neq f}}^{m-1} (1 - \boldsymbol{\phi}_j^r)\right)(-r\boldsymbol{\phi}_f^{r-1})\boldsymbol{\phi}_m^r. \tag{A2}$$

b. When the design field (f) is equal to the number of materials (m):

$$\frac{\partial \mathbf{w}_m^{un}}{\partial \boldsymbol{\phi}_f} = \left(\prod_{j=1}^{m-1} (1 - \phi_j^r)\right) r\boldsymbol{\phi}_m^{r-1}. \tag{A3}$$

iii. For material $m = n_{Mat}$, gradients of un-normalized weight can be computed in tensor form as:

$$\frac{\partial \mathbf{w}_m^{un}}{\partial \boldsymbol{\phi}_f} = \left(\prod_{\substack{j=1 \\ j\neq f}}^{m-1} (1 - \boldsymbol{\phi}_j^r)\right)(-r\boldsymbol{\phi}_f^{r-1}). \tag{A4}$$

Gradient normalization:

$$\frac{\partial \mathbf{w}_m}{\partial \boldsymbol{\phi}_f} = \frac{\left(\frac{\partial \mathbf{w}_m^{un}}{\partial \boldsymbol{\phi}_f}\right)\sum_{m=1}^{n_{Mat}} \mathbf{w}_m^{un} - \mathrm{w}_m^{un}\left(\sum_{m=1}^{n_{Mat}} \frac{\partial \mathbf{w}_m^{un}}{\partial \boldsymbol{\phi}_f}\right)}{\left(\sum_{m=1}^{n_{Mat}} \mathbf{w}_m^{un}\right)^2}. \tag{A5}$$

### 2. Sensitivity calculation for topology optimization for adjacent eigenfrequencies bandgap maximization

Sensitivity is calculated for each candidate material f = 1, 2,…,$n_{Mat}$-1.

i. For the sensitivity of the objective function, the objective function is given as:

$$f_0 = -(\lambda_2 - \lambda_1). \tag{A6}$$

Sensitivity of the objective function:

$$\frac{\partial f_0}{\partial \lambda_1} = 1; \quad \frac{\partial f_0}{\partial \lambda_2} = -1. \tag{A7}$$

Combined sensitivity:

$$\frac{\partial f_0}{\partial \mathbf{x_{mma}}} = [0, \dots, 0,\ 1,\ -1]^T. \tag{A8}$$

ii. For the sensitivity of lower band constraints, for band indices $\mathbf{j} = \mathbf{1}, \dots, \mathbf{j}_{lower}$, the lower band constraints are given as:

$$g_j^{(1)} = \frac{\omega_p^{(j)}}{\lambda_{ref}} - \lambda_1 \leq 0. \tag{A9}$$

For each band j, the p-norm smoothed max across k-points:

$$\omega_p^{(j)} = \omega_{max} = \left(\sum_{n=1}^{N_k} \omega_{\mathbf{k}(n),j}^{p}\right)^{\frac{1}{p}}, \tag{A10}$$

where, $\omega_{\mathbf{k},j}$= frequency at k-point in band j. The element-wise sensitivity w.r.t. design variables is computed as:

$$\frac{\partial g_j^{(1)}}{\partial x_e} = \frac{1}{\lambda_{ref}} \frac{\partial \omega_p^{(j)}}{\partial x_e}, \tag{A11}$$

where

$$\frac{\partial \omega_p^{(j)}}{\partial x_e} = \omega_p^{(j)\,(1-p)} \sum_{n=1}^{N_k} \omega_{\mathbf{k}(n),j}^{(p-1)} \frac{\partial \omega_{\mathbf{k}(n),j}}{\partial x_e}, \tag{A12}$$

$$\frac{\partial \omega_{\mathbf{k},j}}{\partial x_e} = \frac{1}{2\omega_{\mathbf{k},j}} \mathbf{u}_{\mathbf{k},j}^T \left( \frac{\partial \mathbf{K_e}}{\partial x_e} - \omega_{\mathbf{k},j}^2 \frac{\partial \mathbf{M_e}}{\partial x_e} \right) \mathbf{u}_{\mathbf{k},j}. \tag{A13}$$

Gradients of stiffness and mass matrices are computed as:

$$\frac{\partial \mathbf{K_e}}{\partial x_e} = \sum_{m=1}^{n_{Mat}} \left( \mathbf{K_e}^{(m)} \left( \sum_{f=1}^{n_{fields}} \frac{\partial w_{e,m}}{\partial \phi_e^{(f)}} \frac{\partial \phi_e^{(f)}}{\partial x_e^{(f)}} \right) \right), \tag{A14}$$

$$\frac{\partial \mathbf{M_e}}{\partial x_e} = \sum_{m=1}^{n_{Mat}} \left( \mathbf{M_e}^{(m)} \left( \sum_{f=1}^{n_{fields}} \frac{\partial w_{e,m}}{\partial \phi_e^{(f)}} \frac{\partial \phi_e^{(f)}}{\partial x_e^{(f)}} \right) \right), \tag{A15}$$

where $\frac{\partial w_{e,m}}{\partial \phi_e^{(f)}}$ and $\frac{\partial \phi_e^{(f)}}{\partial x_e^{(f)}}$ are the gradient of weight w.r.t. the projected design variable and the gradient of the projected design variable w.r.t. the original variable. Combining all, we have:

$$\frac{\partial \omega_p^{(j)}}{\partial x_e} = \omega_p^{(j)\,(1-p)} \sum_{n=1}^{N_k} \omega_{\mathbf{k}(n),j}^{(p-1)} \ldots \left[ \frac{1}{2\omega_{\mathbf{k}(n),j}} \mathbf{u}_{\mathbf{k}(n),j}^T \left( \frac{\partial \mathbf{K_e}}{\partial x_e} - \omega_{\mathbf{k}(n),j}^2 \frac{\partial \mathbf{M_e}}{\partial x_e} \right) \mathbf{u}_{\mathbf{k},j} \right]. \tag{A16}$$

Writing density filtering of sensitivity in tensor form:

$$\widetilde{\frac{\partial g_j^{(1)}}{\partial \mathbf{x}}} = \mathbf{H} \left( \frac{\frac{\partial g_j^{(1)}}{\partial \mathbf{x}}}{\mathbf{H_s}} \right). \tag{A17}$$

Mapping to reduced variables (IBZ symmetry):

$$\frac{\partial g_j^{(1)}}{\partial \mathbf{x_{red}}} = \frac{1}{\lambda_{ref}} \mathbf{M_{MAP}}^T \widetilde{\frac{\partial g_j^{(1)}}{\partial \mathbf{x}}}. \tag{A18}$$

Sensitivity w.r.t. auxiliary variable $\lambda_1$:

$$\frac{\partial g_j^{(1)}}{\partial \lambda_1} = -1. \tag{A19}$$

Final sensitivity can be then calculated as:

$$\frac{\partial g_j^{(1)}}{\partial \mathbf{x_{mma}}} = \left[ \frac{\partial g_j^{(1)}}{\partial \mathbf{x_{red}}}, \quad \frac{\partial g_j^{(1)}}{\partial \lambda_1}, \quad 0 \right]. \tag{A20}$$

iii. Sensitivity of upper band constraints

For band indices $j = j_{lower} + u$ where $u = 1, \ldots, n_{upper}$, the upper band constraints can be written as:

$$g_u^{(2)} = \lambda_2 - \frac{\omega_{pu}^{(u)}}{\lambda_{refu}} \leq 0. \tag{A21}$$

For each band u, p-norm smoothed mininum across k-points as:

$$\omega_{pu}^{(u)} = \omega_{min} = \left( \sum_{n=1}^{N_k} \omega_{\mathbf{k}(n),j}^{-p} \right)^{-1/p}. \tag{A22}$$

Element-wise sensitivity w.r.t. design variables:

$$\frac{\partial g_u^{(2)}}{\partial x_e} = -\frac{1}{\lambda_{refu}} \frac{\partial \omega_{pu}^{(u)}}{\partial x_e}, \tag{A23}$$

where,

$$\frac{\partial \omega_{pu}^{(u)}}{\partial x_e} = \omega_{pu}^{(u)\,(1+p)} \sum_{n=1}^{N_k} \omega_{\mathbf{k}(n),j}^{-(p+1)} \frac{\partial \omega_{\mathbf{k}(n),j}}{\partial x_e}, \tag{A24}$$

furthermore,

$$\frac{\partial \omega_{\mathbf{k},j}}{\partial x_e} = \frac{1}{2\omega_{\mathbf{k},j}} \mathbf{u}_{\mathbf{k},j}^T \left( \frac{\partial \mathbf{K_e}}{\partial x_e} - \omega_{\mathbf{k},j}^2 \frac{\partial \mathbf{M_e}}{\partial x_e} \right) \mathbf{u}_{\mathbf{k},j}, \tag{A25}$$

where $\frac{\partial \mathbf{K_e}}{\partial x_e}$ and $\frac{\partial \mathbf{M_e}}{\partial x_e}$ are the same as above. We have:

$$\frac{\partial \omega_{pu}^{(u)}}{\partial x_e} = \omega_{pu}^{(u)\,(1+p)} \sum_{n=1}^{N_k} \omega_{\mathbf{k}(n),j}^{-(p+1)} \ldots \left[ \frac{1}{2\omega_{\mathbf{k}(n),j}} \mathbf{u}_{\mathbf{k}(n),j}^T \left( \frac{\partial \mathbf{K_e}}{\partial x_e} - \omega_{\mathbf{k}(n),j}^2 \frac{\partial \mathbf{M_e}}{\partial x_e} \right) \mathbf{u}_{\mathbf{k},j} \right]. \tag{A26}$$

Writing density filtering of sensitivity in tensor form as:

$$\widetilde{\frac{\partial g_j^{(2)}}{\partial \mathbf{x}}} = \mathbf{H} \left( \frac{\frac{\partial g_j^{(2)}}{\partial \mathbf{x}}}{\mathbf{H_s}} \right). \tag{A27}$$

Mapping it to the reduced variables via the IBZ symmetry, we have:

$$\frac{\partial g_u^{(2)}}{\partial \mathbf{x_{red}}} = -\frac{1}{\lambda_{refu}} \mathbf{M_{MAP}}^T \widetilde{\frac{\partial g_j^{(2)}}{\partial \mathbf{x}}}. \tag{A28}$$

Derivative w.r.t. auxiliary variable $\lambda_2$:

$$\frac{\partial g_u^{(2)}}{\partial \lambda_2} = 1. \tag{A29}$$

Final sensitivity can be then computed as:

$$\frac{\partial g_u^{(2)}}{\partial \mathbf{x_{mma}}} = \left[ \frac{\partial g_u^{(2)}}{\partial \mathbf{x_{red}}}, \quad 0, \quad \frac{\partial g_u^{(2)}}{\partial \lambda_2} \right]. \tag{A30}$$

iv. Sensitivity of volume constraints

Sensitivity to volume constraints for each material is calculated from the material's volume fraction. For material $m = 1, \ldots, n_{Mat}$. The volume constraint is:

$$g_m^{(3)} = \frac{v_m}{v_m^{target}} - 1 \leq 0. \tag{A31}$$

Element-wise sensitivity of volume constraint:

$$\frac{\partial g_m^{(3)}}{\partial x_e} = \frac{1}{v_m^{target}} \frac{\partial v_m}{\partial x_e}, \tag{A32}$$

where

$$\frac{\partial v_m}{\partial x_e} = \frac{1}{n_e} \frac{\partial w_{e,m}}{\partial x_e}. \tag{A33}$$

By using the chain rule, we have:

$$\frac{\partial w_{e,m}}{\partial x_e^{(f)}} = \sum_{f=1}^{n_{fields}} \frac{\partial w_{e,m}}{\partial \phi_e^{(f)}} \frac{\partial \phi_e^{(f)}}{\partial x_e^{(f)}}. \tag{A34}$$

Writing density filtering of sensitivity in tensor form:

$$\frac{\widetilde{\partial g_j^{(3)}}}{\partial \mathbf{x}} = \mathbf{H}\left(\frac{\frac{\partial g_m^{(3)}}{\partial \mathbf{x}}}{\mathbf{H_s}}\right). \tag{A35}$$

Symmetry reduction by the IBZ mapping, we have:

$$\frac{\partial g_m^{(3)}}{\partial \mathbf{x}_{\mathbf{red}}} = \frac{1}{v_m^{target}} \mathbf{M_{MAP}}^T \frac{\widetilde{\partial g_j^{(3)}}}{\partial \mathbf{x}}. \tag{A36}$$

Combining all sensitivity fields, we have:

$$\frac{\partial g_m^{(3)}}{\partial \mathbf{x}_{\mathbf{mma}}} = \left[\frac{\partial g_m^{(3)}}{\partial \mathbf{x}_{\mathbf{red}}}, \quad 0, \quad 0\right]. \tag{A37}$$

## 3. Sensitivity for topology optimization for maximum bandgap at targeted frequency range

i. Sensitivity of the objective function. The lower band can be written by using the smoothed maximum across k-points as:

$$\omega_p^{lower} = \left(\sum_{n=1}^{N_k} \omega_{\mathbf{k}(n),j}^{p}\right)^{\frac{1}{p}}. \tag{A38}$$

Upper band (smooth min across k-points):

$$\omega_p^{upper} = \left(\sum_{n=1}^{N_k} \omega_{\mathbf{k}(n),j+1}^{-p}\right)^{-\frac{1}{p}}. \tag{A39}$$

Objective function:

$$f_0 = -\left(\frac{\omega_p^{upper} - \omega_p^{lower}}{\omega_{\text{low, target}}}\right). \tag{A40}$$

Element-wise sensitivity of the objective function:

$$\frac{\partial f_0}{\partial x_e} = -\frac{1}{\omega_{\text{low, target}}}\left(\frac{\partial \omega_p^{upper}}{\partial x_e} - \frac{\partial \omega_p^{lower}}{\partial x_e}\right), \tag{A41}$$

where

$$\frac{\partial \omega_p^{lower}}{\partial x_e} = \omega_p^{lower\,(1-p)} \sum_{n=1}^{N_k} \omega_{\mathbf{k}(n),j}^{p-1} \frac{\partial \omega_{\mathbf{k}(n),j}}{\partial x_e}, \tag{A42}$$

$$\frac{\partial \omega_p^{upper}}{\partial x_e} = \omega_p^{upper\,(1+p)} \sum_{n=1}^{N_k} \omega_{\mathbf{k}(n),j+1}^{-p-1} \frac{\partial \omega_{\mathbf{k},j+1}}{\partial x_e}, \tag{A43}$$

Where the calculation of $\frac{\partial \omega_{\mathbf{k},j}}{\partial x_e}$ is similar to the above. We can write the density filtering of sensitivity in tensor form as:

$$\frac{\widetilde{\partial f_0}}{\partial \mathbf{x}} = \mathbf{H}\left(\frac{\frac{\partial f_0}{\partial \mathbf{x}}}{\mathbf{H_s}}\right). \tag{A44}$$

Mapping to reduced variables via IBZ symmetry as:

$$\frac{\partial f_0}{\partial \mathbf{x}} = \mathbf{M_{MAP}}^T \frac{\widetilde{\partial f_0}}{\partial \mathbf{x}}. \tag{A45}$$

ii. Sensitivity of lower band constraints

$$g_1 = \frac{\omega_p^{lower}}{\omega_{\text{low, target}}} - 1 \le 0. \tag{A46}$$

Sensitivity w.r.t. design variables:

$$\frac{\partial g_1}{\partial x_e} = \frac{1}{\omega_{\text{low, target}}} \frac{\partial \omega_p^{lower}}{\partial x_e}, \tag{A47}$$

where,

$$\frac{\partial g_1}{\partial x_e} = \frac{\omega_p^{lower\,(1-p)}}{\omega_{\text{low, target}}} \sum_{n=1}^{N_k} \omega_{\mathbf{k}(n),j}^{p-1} \frac{\partial \omega_{\mathbf{k}(n),j}}{\partial x_e}. \tag{A48}$$

where the calculation of $\frac{\partial \omega_{\mathbf{k},j}}{\partial x_e}$ is similar to the above.

Writing density filtering of sensitivity in tensor form:

$$\frac{\widetilde{\partial g_1}}{\partial \mathbf{x}} = \mathbf{H}\left(\frac{\frac{\partial g_1}{\partial \mathbf{x}}}{\mathbf{H_s}}\right). \tag{A49}$$

Mapping to reduced variables by IBZ symmetry, we have:

$$\frac{\partial g_1}{\partial \mathbf{x}} = \mathbf{M_{MAP}}^T \frac{\widetilde{\partial g_1}}{\partial \mathbf{x}}. \tag{A50}$$

iii. Sensitivity of upper band constraints

$$g_2 = \frac{\omega_{\text{upp, target}}}{\omega_p^{upper}} - 1 \le 0. \tag{A51}$$

Sensitivity w.r.t. design variables:

$$\frac{\partial g_2}{\partial x_e} = -\frac{\omega_{\text{upp, target}}}{\left(\omega_p^{upper}\right)^2} \frac{\partial \omega_p^{upper}}{\partial x_e}, \tag{A52}$$

$$\frac{\partial g_2}{\partial x_e} = -\omega_{upp}\left(\omega_p^{upper}\right)^{p-1} \sum_{n=1}^{N_k} \omega_{\mathbf{k}(n),j+1}^{-p-1} \frac{\partial \omega_{\mathbf{k}(n),j+1}}{\partial x_e}, \tag{A53}$$

Writing density filtering of sensitivity in tensor form:

$$\frac{\widetilde{\partial g_2}}{\partial \mathbf{x}} = \mathbf{H}\left(\frac{\frac{\partial g_2}{\partial \mathbf{x}}}{\mathbf{H_s}}\right). \tag{A54}$$

Mapping to reduced variables by IBZ symmetry:

$$\frac{\partial g_2}{\partial \mathbf{x}} = \mathbf{M_{MAP}}^T \frac{\widetilde{\partial g_2}}{\partial \mathbf{x}}. \tag{A55}$$

iv. Sensitivity of volume constraints

The sensitivity to volume constraints in topology optimization for maximum bandgap at the targeted frequency range is the same as in topology optimization for adjacent eigenfrequencies bandgap maximization.

## APPENDIX B: $\mathbf{M_{MAP}}$ matrix

Let the full design domain have the number of elements as $n_e = n_{elx} \times n_{ely}$. Then, half size is obtained as $n_{ex} = \frac{n_{elx}}{2}, n_{ey} = \frac{n_{ely}}{2}$. The number of reduced variables by triangular IBZ is as:

$$n_{red} = \frac{n_{ex}(n_{ex} + 1)}{2}. \tag{B1}$$

Mapping matrix:

$$\mathbf{x_{full}} = \mathbf{M_{MAP}} * \mathbf{x_{red}} \tag{B2}$$

Reduced index with triangular indexing:

$$k(i,j) = \sum_{r=1}^{j-1} (n_{ex} - r + 1) + (i - j + 1), \tag{B3}$$

for $j = 1, \dots, n_{ey}$; $i = j, \dots, n_{ex}$.

For each reduced element $(i, j)$, we generate 8 symmetric copies:

- Original:

$$e_1 = (j-1)n_{elx} + i. \quad \text{(B4)}$$

- Mirror in x-direction:

$$e_2 = (j-1)n_{elx} + (n_{elx} - i + 1). \quad \text{(B5)}$$

- Mirror in y-direction:

$$e_3 = \left(n_{ely} - j\right)n_{elx} + i. \quad \text{(B6)}$$

- Mirror in both x and y:

$$e_4 = \left(n_{ely} - j\right)n_{elx} + (n_{elx} - i + 1). \quad \text{(B7)}$$

- Diagonal swap (transpose symmetry):

$$e_5 = (i-1)n_{elx} + j. \quad \text{(B8)}$$

- Diagonal + y reflection:

$$e_6 = (i-1)n_{elx} + \left(n_{ely} - j + 1\right). \quad \text{(B9)}$$

- Diagonal + x reflection:

$$e_7 = (n_{elx} - i)n_{elx} + j. \quad \text{(B10)}$$

- Diagonal + x & y reflection:

$$e_8 = (n_{elx} - i)n_{elx} + \left(n_{ely} - j + 1\right). \quad \text{(B11)}$$

For each symmetric index $e_m$, we assign:

$$\mathbf{M_{MAP}}(e,k) = \begin{cases} 1, & \text{if } e \in \{e_1, \dots, e_8\} \\ 0, & \text{otherwise} \end{cases}. \quad \text{(B12)}$$

We have the mapping relation function as:

$$\mathbf{x_{full}}(e) = \sum_{k=1}^{n_{red}} \mathbf{M_{MAP}}(e,k)\ \mathbf{x_{red}}(k). \quad \text{(B13)}$$